\documentclass[preprint2]{aastex62}

\usepackage{natbib} 
\usepackage{graphicx}
\usepackage{enumitem}

\usepackage{makecell}
\usepackage{longtable}

\shorttitle{SAFARI: Searching Asteroids For Activity Revealing Indicators}
\shortauthors{Chandler et. al}
\newcommand{\allthumbs}{15,600}
\newcommand{\uniquethumbs}{11,703}

\newcommand{\fitscount}{35,640}
\newcommand{\gifcount}{3,029}

\newcommand{\sublimate}{$\wr\hspace{-0.5mm}\wr\hspace{-0.5mm}\wr$}

\usepackage{calc}
\newlength{\okinalen}
\newcommand{\omuamua}{\hbox to.666\okinalen{\hss`\hss}Oumuamua}

\newcounter{rownumber}
\newcommand{\rlabel}[1]{\refstepcounter{rownumber}\label{#1}}
\newcounter{tabfoot}
\newcommand{\flabel}[1]{\refstepcounter{tabfoot}\label{#1}}

\begin{document}

\title{SAFARI: Searching Asteroids For Activity Revealing Indicators

}

\author[0000-0001-7335-1715]{Colin Orion Chandler}
\affiliation{Department of Physics \& Astronomy, Northern Arizona University, PO Box 6010, Flagstaff, AZ 86011, USA; orion@nau.edu}

\author[0000-0002-0212-4563]{Anthony M. Curtis}
\affiliation{Department of Physics, University of South Florida ISA 2019, Tampa, FL 33620, USA}

\author[0000-0002-8132-778X]{Michael Mommert}
\affiliation{Department of Physics \& Astronomy, Northern Arizona University, PO Box 6010, Flagstaff, AZ 86011, USA; orion@nau.edu}

\author[0000-0003-3145-8682]{Scott S. Sheppard}
\affiliation{Department of Terrestrial Magnetism, Carnegie Institution for Science, 5241 Broad Branch Road. NW, Washington, DC 20015, USA}

\author[0000-0001-9859-0894]{Chadwick A. Trujillo}
\affiliation{Department of Physics \& Astronomy, Northern Arizona University, PO Box 6010, Flagstaff, AZ 86011, USA; orion@nau.edu}


\begin{abstract}
\label{Abstract}
Active asteroids behave dynamically like asteroids but display comet-like comae. These objects are poorly understood, with only about 30 identified to date. We have conducted one of the deepest systematic searches for asteroid activity by making use of deep images from the Dark Energy Camera (DECam) ideally suited to the task. We looked for activity indicators amongst \uniquethumbs{} unique asteroids extracted from \fitscount{} images. We detected three previously-identified active asteroids ((62412), (1) Ceres and (779) Nina), though only (62412) showed signs of activity. Our activity occurrence rate of 1 in \uniquethumbs{} is consistent with the prevailing 1 in 10,000 activity occurrence rate estimate. Our proof of concept demonstrates 1) our novel informatics approach can locate active asteroids and 2) DECam data are well-suited to the search for active asteroids.
\end{abstract}


\section{Introduction}
\label{introduction}

\begin{figure}
  \centering
	\includegraphics[width=1.0\linewidth]{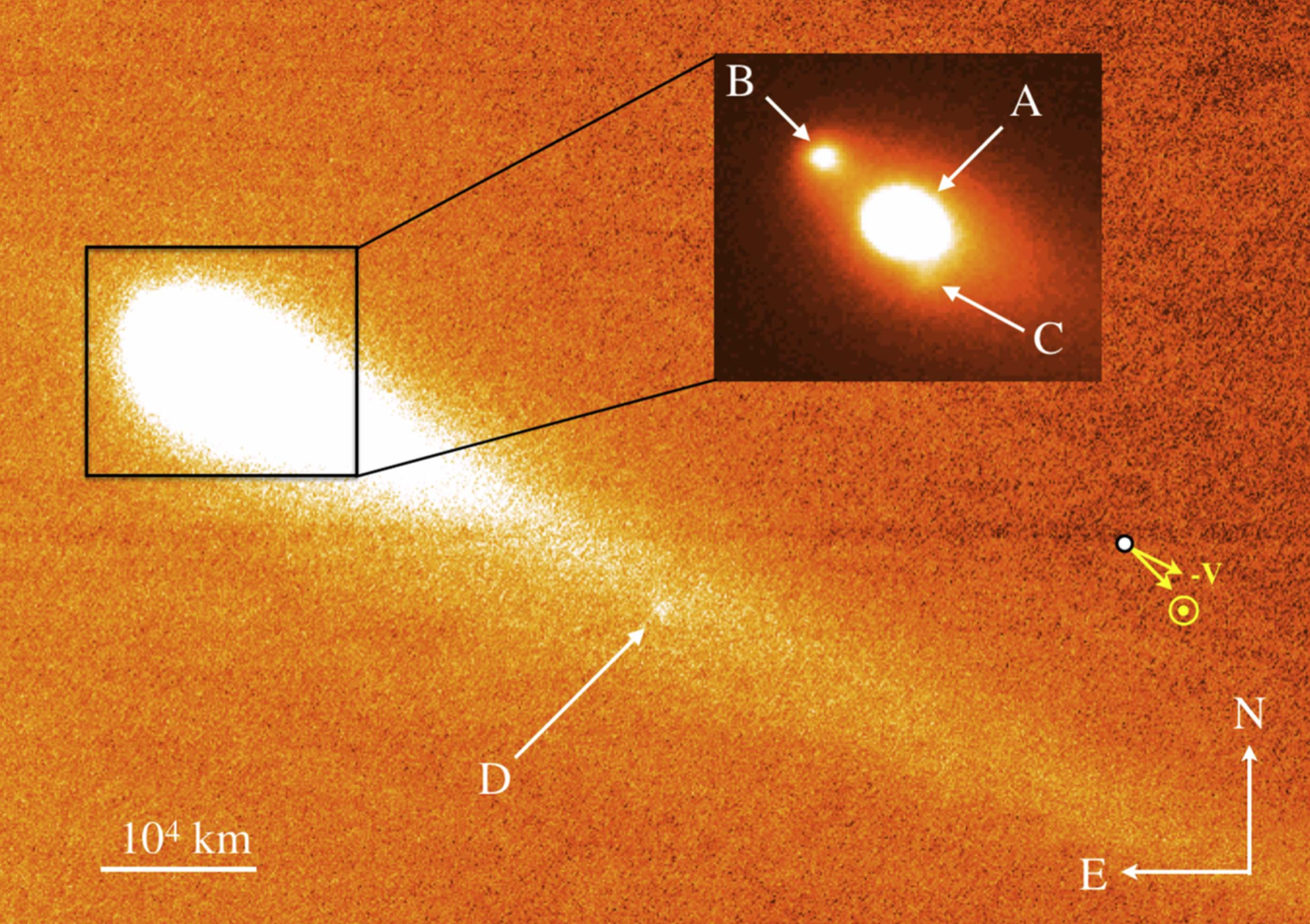}
  \caption{Active asteroid P2013/R3 was imaged in October 2013 while undergoing a breakup (into components A-D) likely caused by rotational instability. The antisolar and negative heliocentric velocity vector arrows are labeled $\odot$ and $-V$, respectively. Reprinted Figure 2 of \cite{Jewitt:2017fa}.}
  \label{ExampleAsteroid}
\end{figure}

\begin{deluxetable*}{lhccrrclcrrrhhhlhchh}
\tablecaption{The Active Asteroids (1 of 2) \label{Table:TheAAs}}
\tabletypesize{\footnotesize}
\tablehead{\colhead{Asteroid Name} & \nocolhead{Family} & \colhead{$a$} & \colhead{$e$} & \colhead{$i$} & \colhead{Orb.} & \colhead{$T_J$} &\colhead{$P$} & \colhead{$q$} & \colhead{\flabel{rSolDist}$R^\mathrm{\ref{rSolDist}}$} & \colhead{\flabel{rTrueAn} $f^\mathrm{\ref{rTrueAn}}$} & \colhead{\flabel{rPP}$\mathrm{\%}_\mathrm{peri}^\mathrm{\ref{rPP}}$} & \nocolhead{\flabel{rfirstact}$1^\mathrm{st}$Act$^\mathrm{\ref{rfirstact}}$} & \nocolhead{\flabel{rfac}Facility$^\mathrm{\ref{rfac}}$} & \nocolhead{Method} & \colhead{\flabel{ract}Act.$^\mathrm{\ref{ract}}$} & \nocolhead{\flabel{rLast}Last$^\mathrm{\ref{rLast}}$} & \colhead{Cause} & \nocolhead{Visit} & \nocolhead{\flabel{rrefs}Refs$^\mathrm{\ref{rrefs}}$}\\ & & (au) & & ($^\circ$) & & & (yr) & (au) & (au) & ($^\circ$) & (\%) & (yr) & & & (N) & (yr) & &}
\tablecolumns{19}
\startdata
\rlabel{Ceres}		(1) Ceres 			& None 		& 2.77 	& 0.08 	& 10.6 	& MB 		& 3.310 & 4.6 	& 2.60 & 2.72 & 279.3 & 62 	& 2014 & Herschel 	& Spec.  & 3+ 	 & 2017 	& \sublimate{}\ , $\bigwedge$ 					& Yes	& [\ref{Ceres}]\\
\rlabel{Adeona}		(145) Adeona  		& Adeona		& 2.67 	& 0.14 	& 12.6 	& MB 		& 3.331 & 2.28 	& 2.29 & 2.69 & 258.8 & 47 	& 2017 & Terksol 	& Spec.  & 2 	 & 2016 	& \sublimate 									& No 	& [\ref{Adeona}]\\
\rlabel{Constantia}	(315) Constantia 	& Flora		& 2.24 	& 0.17 	&  2.4 	& MB 		& 3.614 & 3.36 	& 1.86 & 1.94 & 315.9 & 92 	& 2013 & MPCAT 		& Phot.  & \flabel{rCik}0$^\mathrm{\ref{rCik}}$ & 2013 	& (?) 											& No 	& [\ref{Constantia}]\\
\rlabel{Griseldis}	(493) Griseldis 		& Eunomia	& 3.12 	& 0.18 	& 15.2  & OMB 		& 3.140 & 5.5 	& 2.57 & 3.34 & 122.4 & 31 	& 2015 & Subaru 	& Visual & 1 	 & 2015 	& $\leftrightsquigarrow$ 						& No 	& [\ref{Griseldis}]\\
\rlabel{Scheila}	(596) Scheila 		& None		& 2.93 	& 0.16 	& 14.7 	& OMB 		& 3.209 & 5.01 	& 2.45 & 3.11 & 239.2 & 90 	& 2010 & CSS 		& Visual & 1 	 & 2010 	& $\leftrightsquigarrow$ 						& No 	& [\ref{Scheila}]\\
\rlabel{Interamnia}	(704) Interamnia 	& None		& 3.06 	& 0.15 	& 17.3 	& MB 		& 3.148 & 5.35 	& 2.59 & 2.62 &  19.6 & 97 	& 2017 & Terksol 	& Spec.  & 1 	 & 2012 	& \sublimate 									& No 	& [\ref{Interamnia}]\\
\rlabel{Nina}		(779) Nina 			& $\cdots$	& 2.66 	& 0.23 	& 14.6 	& MB 		& 3.302 & 4.35 	& 2.06 & 2.15 &  36.9 & 93 	& 2017 & Terksol 	& Spec.  & 1 	 & 2012 	& \sublimate 									& No 	& [\ref{Nina}]\\
\rlabel{Ingrid}		(1026) Ingrid 		& Flora 		& 2.25 	& 0.18 	&  5.4 	& MB 		& 3.597 & 3.38 	& 1.85 & 2.23 &  97.5 & 16 	& 2013 & MPCAT 		& Phot.  & 0$^\mathrm{\ref{rCik}}$ & 2013 	& (?) 											& No 	& [\ref{Ingrid}]\\
\rlabel{Beira}		(1474) Beira 		& $\cdots$	& 2.73 	& 0.49 	& 26.7 	& Mars 		& 3.033 & 4.52 	& 1.39 & 1.57 & 310.9 & 93 	& 2017 & Terksol 	& Spec.  & 1 	 & 2012 	&  \sublimate 									& No 	& [\ref{Beira}]\\
\rlabel{Oljato}		(2201) Oljato 		& $\cdots$ 	& 2.17 	& 0.71 	&  2.5	& Apollo 	& 3.298 & 3.21 	& 0.62 & 0.88 &  73.1 & 92 	& 1984 & Pioneer 	& Mag.   & 1 	 & 1984 	& (?) 											& No 	& [\ref{Oljato}]\\
\rlabel{Phaethon}	(3200) Phaethon 		& Pallas 	& 1.27 	& 0.89 	& 22.2 	& Apollo 	& 4.510 & 1.43 	& 0.14 & 0.14 &   5.1 & 87 	& 2009 & STEREO 	& Visual & 3 	 & 2017 	& $\bigodot$ 									& Yes 	& [\ref{Phaethon}]\\
\rlabel{DonQuixote}	(3552) Don Quixote 	& $\cdots$	& 4.26 	& 0.71 	& 31.1 	& Amor 		& 2.315 & 8.78 	& 1.24 & 1.23 & 343.6 & 100	& 2009 & Spitzer 	& Visual & 2 	 & 2018 	& \sublimate,(?) 								& No 	& [\ref{DonQuixote}]\\
\rlabel{Aduatiques}	(3646) Aduatiques 	& $\cdots$	& 2.75 	& 0.11 	&  0.6 	& MB 		& 3.336 & 4.57 	& 2.46 & 2.56 & 309.0 & 90 	& 2013 & MPCAT 		& Phot.  & 0$^\mathrm{\ref{rCik}}$ & 2013 	& (?) 											& No 	& [\ref{Aduatiques}]\\
\rlabel{WilHar}		(4015) Wil.-Har. 	& $\cdots$	& 2.63 	& 0.63 	&  2.8 	& Apollo 	& 3.082 & 4.26 	& 0.97 & 1.17 &  51.0 & 95 	& 1949 & Palomar 	& Visual & \flabel{rWH}2$^\mathrm{\ref{rWH}}$ & 1979$^\mathrm{\ref{rWH}}$	& \sublimate{}\ , (?) 							& No 	& [\ref{WilHar}]\\
\rlabel{24684}		(24684) 1990 EU4 	& $\cdots$	& 2.32 	& 0.08 	&  3.9	& MB 		& 3.572 & 3.53 	& 2.13 & 2.28 & 277.9 & 77 	& 2013 & MPCAT 		& Phot.  & 0$^\mathrm{\ref{rCik}}$ & 2013 	& (?) 											& No 	& [\ref{24684}]\\
\rlabel{35101}		(35101) 1991 PL16 	& Eunomia	& 2.60 	& 0.18 	& 12.3 	& MB 		& 3.365 & 4.17 	& 2.12 & 2.86 & 227.0 & 21 	& 2013 & MPCAT 		& Phot.  & 0$^\mathrm{\ref{rCik}}$ & 2013 	& (?) 											& No 	& [\ref{35101}]\\
\rlabel{62412}		(62412) 				& Hygiea		& 3.15 	& 0.08 	&  4.7	& OMB 		& 3.197 & 5.6 	& 2.90 & 3.06 &  74.5 & 68 	& 2015 & DECam 		& Visual & 1 	 & 2014 	& $\circlearrowleft$ 							& No 	& [\ref{62412}]\\
\rlabel{Ryugu}		(162173) Ryugu 		& Clarissa	& 1.19 	& 0.19 	&  5.9	& Apollo 	& 5.308 & 1.3 	& 0.96 & 1.08 & 288.4 & 8 	& 2017 & MMT 		& Spec.	 & 1 	 & 2017 	& \sublimate 									& Yes 	& [\ref{Ryugu}]\\
\rlabel{GO98}		(457175) 			& Hilda		& 3.96 	& 0.28 	& 15.6 	& OMB 		& 2.926 & 7.89 	& 2.85 & 3.28 &  66.0 & 81 	& 2017 & CSS 		& Visual & 1 	 & 2017 	& (?) 											& No 	& [\ref{GO98}]\\
\rlabel{ElstPizarro} 133P/Elst-Pizarro 	& Themis		& 3.16 	& 0.16 	&  1.4	& OMB 		& 3.184 & 5.63 	& 2.66 & 2.65 &  21.7 & 100	& 1996 & ESO 		& Visual & 4 	 & 2014 	& \sublimate 									& No 	& [\ref{ElstPizarro}]\\	
\rlabel{176P}		176P/LINEAR 			& Themis		& 3.20 	& 0.19 	&  0.2	& OMB 		& 3.166 & 5.71 	& 2.58 & 2.59 &  10.1 & 1 	& 2009 & HTP 		& Visual & 1 	 & 2011 	& (?) 											& No 	& [\ref{176P}]\\
\rlabel{233P}		233P/La Sagra 		& $\cdots$	& 3.04 	& 0.41 	& 11.3 	& Encke 	& 3.081 & 5.28 	& 1.78 & 2.01 & 309.1 & 91 	& 2009 & LSSS 		& Visual & 1 	 & 2009 	& (?) 											& No 	& [\ref{233P}]\\
\rlabel{238P}		238P/Read 			& Gorchakov	& 3.16 	& 0.25 	& 1.3 	& OMB 		& 3.154 & 5.64 	& 2.37 & 2.42 &  26.5 & 97 	& 2005 & SW 		& Visual & 3 	 & 2016 	& \sublimate 									& No 	& [\ref{238P}]\\
\rlabel{259P}		259P/Garradd 		& $\cdots$	& 2.73 	& 0.34 	& 15.9 	& MMB 		& 3.217 & 4.51 	& 1.81 & 1.85 &  27.6 & 99 	& 2008 & SS 		& Visual & 2 	 & 2017 	& \sublimate 									& No 	& [\ref{259P}]\\
\rlabel{288P}		288P (300163) 		& Themis		& 3.05 	& 0.20 	&  3.2 	& OMB 		& 3.204 & 5.32 	& 2.44 & 2.45 &  12.2 & 99 	& 2011 & PS 		& Visual & 2 	 & 2017 	& \sublimate 									& No 	& [\ref{288P}]\\
\rlabel{311P}		311P/PS 				& Behrens 	& 2.19 	& 0.12 	&  5.0	& IMB 		& 3.661 & 3.24 	& 1.94 & 2.15 & 272.8 & 58 	& 2013 & PS 		& Visual & 2 	 & 2014 	& $\circlearrowleft$\ , \large{\textbf{:}} 		& No 	& [\ref{311P}]\\
\rlabel{313P}		313P/Gibbs 			& Lixiaohua & 3.16 	& 0.24 	& 11.0 	& Encke 	& 3.132 & 5.62 	& 2.42 & 2.40 &   8.0 & 100	& 2014 & CSS 		& Visual & 2 	 & 2015 	& \sublimate 									& No 	& [\ref{313P}]\\
\rlabel{324P}		324P/La Sagra 		& Alauda		& 3.10 	& 0.15 	& 21.4 	& OMB 		& 3.100 & 5.45 	& 2.62 & 2.64 &  20.0 & 98 	& 2011 & LSSS 		& Visual & 2 	 & 2015 	& \sublimate 									& No 	& [\ref{324P}]\\
\rlabel{331P}		331P/Gibbs 			& Gibbs		& 3.00 	& 0.04 	&  9.7	& OMB 		& 3.229 & 5.21 	& 2.88 & 3.10 & 140.4 & 11 	& 2012 & CSS 		& Visual & 2 	 & 2014 	& $\leftrightsquigarrow$, $\circlearrowleft$ 	& No 	& [\ref{331P}]\\
\rlabel{348P}		348P/PS 				& $\cdots$	& 3.17 	& 0.30 	& 17.6 	& OMB 		& 3.062 & 5.63 	& 2.18 & 2.51 &  60.8 & 83 	& 2017 & PS 		& Visual & 1 	 & 2017 	& (?) 											& No 	& [\ref{348P}]\\
\rlabel{354P}		354P/LINEAR 			& Baptistina& 2.29 	& 0.12 	&  5.3	& OMB 		& 3.583 & 3.47 	& 2.00 & 2.01 &  12.2 & 99 	& 2010 & LINEAR 	& Visual & 1 	 & 2017 	& $\circlearrowleft$,$\circledast$ 				& No 	& [\ref{354P}]\\
\rlabel{358P}		358P 				& Lixiaohua	& 3.15 	& 0.24 	& 11.1 	& Encke 	& 3.135 & 5.59 	& 2.39 & 2.42 &   7.5 & 99 	& 2012 & PS 		& Visual & 2 	 & 2017 	& \sublimate{}\ , (?) 							& No 	& [\ref{358P}]\\
\rlabel{P2013R3}	P/2013 R3 			& Mandragora& 3.03 	& 0.27 	& 0.9 	& OMB 		& 3.184 & 5.28 	& 2.20 & 2.22 &  14.0 & 99 	& 2013 & PS 		& Visual & 1 	 & 2013 	& $\circlearrowleft$, \sublimate{} 				& No 	& [\ref{P2013R3}]\\
\rlabel{P2015X6}	P/2015 X6 			& Aeolia		& 2.75	& 0.17 	&  4.6	& MMB 		& 3.318 & 4.57 	& 2.28 & 2.64 & 274.5 & 62 	& 2015 & PS 		& Visual & 1 	 & 2015 	& $\circlearrowleft$ 							& No 	& [\ref{P2015X6}]\\
\rlabel{P2016G1}	P/2016 G1 			& Adeona		& 2.58	& 0.21	& 11.0	& MMB 		& 3.367 & 4.15 	& 2.04 & 2.52 & 264.7 & 56 	& 2016 & PS 		& Visual & 1 	 & 2016 	& $\leftrightsquigarrow$ 						& No 	& [\ref{P2016G1}]\\
\rlabel{P2016J1}	P/2016 J1 			& Theobalda	& 3.17 	& 0.23 	& 14.3 	& OMB 		& 3.113 & 5.65 	& 2.45 & 2.46 & 345.9 & 99 	& 2016 & PS 		& Visual & 1 	 & 2016 	& $\circlearrowleft$, \sublimate{} 				& No 	& [\ref{P2016J1}]\\
\enddata
\tablenotetext{}{}
	Orbital parameters retrieved from the Minor Planet Center and JPL Horizons. %
	$T_\mathrm{J}$:Tisserand parameter with respect to Jupiter; %
	$P$:Orbital Period; %
	$d_\mathrm{peri}$:Perihelion distance; %
	$\mathrm{\%}_\mathrm{peri}$:How close (\%) an object was to perihelion at activity discovery time (see text); %
	
	$^\mathrm{\ref{rSolDist}}$Heliocentric discovery distance. %
	$f^\mathrm{\ref{rTrueAn}}$True anomaly. %
	$^\mathrm{\ref{rfirstact}}$Year activity discovered. %
	$^\mathrm{\ref{rfac}}$Facility originally reporting activity. 
	$^\mathrm{\ref{ract}}$Number of times object reported active. 
	$^\mathrm{\ref{rLast}}$As of January 2018 submission. %
	$^\mathrm{\ref{rrefs}}$Object-specific references in Appendix. %
	$^\mathrm{\ref{rCik}}$Authors declare object a candidate (activity not yet confirmed). %
	$^\mathrm{\ref{rWH}}$\cite{Ferrin:2012gm} argue (4015) was also active in 1992, 1996, 2008, and 2009-2010. %
	
	\sublimate{}\hspace{0.5mm}Sublimation; %
	$\circlearrowleft$\hspace{0.25mm}Rotational Breakup; %
	$\leftrightsquigarrow$\hspace{0.25mm}Impact; %
	$\bigwedge$\hspace{0.25mm}Cryovolcanism; %
	\protect{\large{\textbf{:}}}\hspace{0.25mm}Binary; %
	$\bigodot$\hspace{0.25mm}Thermal Fracture; %
	$\circledast$\hspace{0.25mm}Dust Model; %
	(?)\hspace{0.25mm}Unknown %
\end{deluxetable*}


\setcounter{table}{0} 

\begin{deluxetable*}{llhhhhhhhhhcllhchcc}
\tablecaption{The Active Asteroids (2 of 2) \label{Table:TheAAs2}}
\tabletypesize{\footnotesize}
\tablehead{\colhead{Asteroid Name} & \colhead{Family} & \nocolhead{$a$} & \nocolhead{$e$} & \nocolhead{$i$} & \nocolhead{Orb.} & \nocolhead{$T_J$} &\nocolhead{$P$} & \nocolhead{$q$} & \nocolhead{\flabel{rSolDist}$R^\mathrm{\ref{rSolDist}}$} & \nocolhead{\flabel{rPP}$\mathrm{\%}_\mathrm{peri}^\mathrm{\ref{rPP}}$} & \colhead{\flabel{rfirstact}$1^\mathrm{st}$Act$^\mathrm{\ref{rfirstact}}$} & \colhead{\flabel{rfac}Facility$^\mathrm{\ref{rfac}}$} & \colhead{Method} & \nocolhead{\flabel{ract}Act.$^\mathrm{\ref{ract}}$} & \colhead{\flabel{rLast}Last$^\mathrm{\ref{rLast}}$} & \nocolhead{Cause} & \colhead{Visit} & \colhead{\flabel{rrefs}Refs$^\mathrm{\ref{rrefs}}$}\\ & & (au) & & ($^\circ$) & & & (yr) & (au) & (au) & (\%) & (yr) & & & (N) & (yr) & &}
\tablecolumns{19}
\startdata
(1) Ceres 			& None 		& 2.77 	& 0.08 	& 10.6 	& MB 		& 3.310 & 4.6 	& 2.60 & 2.72 & 62 	& 2014 & Herschel 	& Spec.  & 3+ 	 & 2017 	& 					& Yes	& [\ref{Ceres}]\\
(145) Adeona  		& Adeona		& 2.67 	& 0.14 	& 12.6 	& MB 		& 3.331 & 2.28 	& 2.29 & 2.69 & 47 	& 2017 & Terksol 	& Spec.  & 2 	 & 2016 	& & No 	& [\ref{Adeona}]\\
(315) Constantia 	& Flora		& 2.24 	& 0.17 	&  2.4 	& MB 		& 3.614 & 3.36 	& 1.86 & 1.94 & 92 	& 2013 & MPCAT 		& Phot.  & \flabel{rCik}0$^\mathrm{\ref{rCik}}$ & 2013 	& & No 	& [\ref{Constantia}]\\
(493) Griseldis 		& Eunomia	& 3.12 	& 0.18 	& 15.2  & OMB 		& 3.140 & 5.5 	& 2.57 & 3.34 & 31 	& 2015 & Subaru 	& Visual & 1 	 & 2015 	& & No 	& [\ref{Griseldis}]\\
(596) Scheila 		& None		& 2.93 	& 0.16 	& 14.7 	& OMB 		& 3.209 & 5.01 	& 2.45 & 3.11 & 90 	& 2010 & CSS 		& Visual & 1 	 & 2010 	& & No 	& [\ref{Scheila}]\\
(704) Interamnia 	& None		& 3.06 	& 0.15 	& 17.3 	& MB 		& 3.148 & 5.35 	& 2.59 & 2.62 & 97 	& 2017 & Terksol 	& Spec.  & 1 	 & 2012 	& & No 	& [\ref{Interamnia}]\\
(779) Nina 			& $\cdots$	& 2.66 	& 0.23 	& 14.6 	& MB 		& 3.302 & 4.35 	& 2.06 & 2.15 & 93 	& 2017 & Terksol 	& Spec.  & 1 	 & 2012 	&	& No 	& [\ref{Nina}]\\
(1026) Ingrid 		& Flora 		& 2.25 	& 0.18 	&  5.4 	& MB 		& 3.597 & 3.38 	& 1.85 & 2.23 & 16 	& 2013 & MPCAT 		& Phot.  & 0$^\mathrm{\ref{rCik}}$ & 2013 	&										& No 	& [\ref{Ingrid}]\\
(1474) Beira 		& $\cdots$	& 2.73 	& 0.49 	& 26.7 	& Mars 		& 3.033 & 4.52 	& 1.39 & 1.57 & 93 	& 2017 & Terksol 	& Spec.  & 1 	 & 2012 	&  & No 	& [\ref{Beira}]\\
(2201) Oljato 		& $\cdots$ 	& 2.17 	& 0.71 	&  2.5	& Apollo 	& 3.298 & 3.21 	& 0.62 & 0.88 & 92 	& 1984 & Pioneer 	& Mag.   & 1 	 & 1984 	& 	& No 	& [\ref{Oljato}]\\
(3200) Phaethon 		& Pallas 	& 1.27 	& 0.89 	& 22.2 	& Apollo 	& 4.510 & 1.43 	& 0.14 & 0.14 & 87 	& 2009 & STEREO 	& Visual & 3 	 & 2017 	&	& Yes 	& [\ref{Phaethon}]\\
(3552) Don Quixote 	& $\cdots$	& 4.26 	& 0.71 	& 31.1 	& Amor 		& 2.315 & 8.78 	& 1.24 & 1.23 & 100	& 2009 & Spitzer 	& Visual & 2 	 & 2018 	&	& No 	& [\ref{DonQuixote}]\\
(3646) Aduatiques 	& $\cdots$	& 2.75 	& 0.11 	&  0.6 	& MB 		& 3.336 & 4.57 	& 2.46 & 2.56 & 90 	& 2013 & MPCAT 		& Phot.  & 0$^\mathrm{\ref{rCik}}$ & 2013 	& & No 	& [\ref{Aduatiques}]\\
(4015) Wil.-Har. 	& $\cdots$	& 2.63 	& 0.63 	&  2.8 	& Apollo 	& 3.082 & 4.26 	& 0.97 & 1.17 & 95 	& 1949 & Palomar 	& Visual & \flabel{rWH}2$^\mathrm{\ref{rWH}}$ & 1979$^\mathrm{\ref{rWH}}$	&	& No 	& [\ref{WilHar}]\\
(24684) 1990 EU4 	& $\cdots$	& 2.32 	& 0.08 	&  3.9	& MB 		& 3.572 & 3.53 	& 2.13 & 2.28 & 77 	& 2013 & MPCAT 		& Phot.  & 0$^\mathrm{\ref{rCik}}$ & 2013 	& & No 	& [\ref{24684}]\\
(35101) 1991 PL16 	& Eunomia	& 2.60 	& 0.18 	& 12.3 	& MB 		& 3.365 & 4.17 	& 2.12 & 2.86 & 21 	& 2013 & MPCAT 		& Phot.  & 0$^\mathrm{\ref{rCik}}$ & 2013 	& & No 	& [\ref{35101}]\\
(62412) 				& Hygiea		& 3.15 	& 0.08 	&  4.7	& OMB 		& 3.197 & 5.6 	& 2.90 & 3.06 & 68 	& 2015 & DECam 		& Visual & 1 	 & 2014 	&	& No 	& [\ref{62412}]\\
(162173) Ryugu 		& Clarissa	& 1.19 	& 0.19 	&  5.9	& Apollo 	& 5.308 & 1.3 	& 0.96 & 1.08 & 8 	& 2017 & MMT 		& Spec.	 & 1 	 & 2017 	&	& Yes 	& [\ref{Ryugu}]\\
(457175) 			& Hilda		& 3.96 	& 0.28 	& 15.6 	& OMB 		& 2.926 & 7.89 	& 2.85 & 3.28 & 81 	& 2017 & CSS 		& Visual & 1 	 & 2017 	& & No 	& [\ref{GO98}]\\
133P/Elst-Pizarro 	& Themis		& 3.16 	& 0.16 	&  1.4	& OMB 		& 3.184 & 5.63 	& 2.66 & 2.65 & 100	& 1996 & ESO 		& Visual & 4 	 & 2014 	& & No 	& [\ref{ElstPizarro}]\\	
176P/LINEAR 			& Themis		& 3.20 	& 0.19 	&  0.2	& OMB 		& 3.166 & 5.71 	& 2.58 & 2.59 & 1 	& 2009 & HTP 		& Visual & 1 	 & 2011 	& & No 	& [\ref{176P}]\\
233P/La Sagra 		& $\cdots$	& 3.04 	& 0.41 	& 11.3 	& Encke 	& 3.081 & 5.28 	& 1.78 & 2.01 & 91 	& 2009 & LSSS 		& Visual & 1 	 & 2009 	& & No 	& [\ref{233P}]\\
238P/Read 			& Gorchakov	& 3.16 	& 0.25 	& 1.3 	& OMB 		& 3.154 & 5.64 	& 2.37 & 2.42 & 97 	& 2005 & SW 		& Visual & 3 	 & 2016 	& & No 	& [\ref{238P}]\\
259P/Garradd 		& $\cdots$	& 2.73 	& 0.34 	& 15.9 	& MMB 		& 3.217 & 4.51 	& 1.81 & 1.85 & 99 	& 2008 & SS 		& Visual & 2 	 & 2017 	& & No 	& [\ref{259P}]\\
288P (300163) 		& Themis		& 3.05 	& 0.20 	&  3.2 	& OMB 		& 3.204 & 5.32 	& 2.44 & 2.45 & 99 	& 2011 & PS 		& Visual & 2 	 & 2017 	& & No 	& [\ref{288P}]\\
311P/PS 				& Behrens 	& 2.19 	& 0.12 	&  5.0	& IMB 		& 3.661 & 3.24 	& 1.94 & 2.15 & 58 	& 2013 & PS 		& Visual & 2 	 & 2014 	& & No 	& [\ref{311P}]\\
313P/Gibbs 			& Lixiaohua & 3.16 	& 0.24 	& 11.0 	& Encke 	& 3.132 & 5.62 	& 2.42 & 2.40 & 100	& 2014 & CSS 		& Visual & 2 	 & 2015 	& & No 	& [\ref{313P}]\\
324P/La Sagra 		& Alauda		& 3.10 	& 0.15 	& 21.4 	& OMB 		& 3.100 & 5.45 	& 2.62 & 2.64 & 98 	& 2011 & LSSS 		& Visual & 2 	 & 2015 	& & No 	& [\ref{324P}]\\
331P/Gibbs 			& Gibbs		& 3.00 	& 0.04 	&  9.7	& OMB 		& 3.229 & 5.21 	& 2.88 & 3.10 & 11 	& 2012 & CSS 		& Visual & 2 	 & 2014 	& & No 	& [\ref{331P}]\\
348P/PS 				& $\cdots$	& 3.17 	& 0.30 	& 17.6 	& OMB 		& 3.062 & 5.63 	& 2.18 & 2.51 & 83 	& 2017 & PS 		& Visual & 1 	 & 2017 	& 	& No 	& [\ref{348P}]\\
354P/LINEAR 			& Baptistina& 2.29 	& 0.12 	&  5.3	& OMB 		& 3.583 & 3.47 	& 2.00 & 2.01 & 99 	& 2010 & LINEAR 	& Visual & 1 	 & 2017 	&	& No 	& [\ref{354P}]\\
358P 				& Lixiaohua	& 3.15 	& 0.24 	& 11.1 	& Encke 	& 3.135 & 5.59 	& 2.39 & 2.42 & 99 	& 2012 & PS 		& Visual & 2 	 & 2017 	& & No 	& [\ref{358P}]\\
P/2013 R3 			& Mandragora& 3.03 	& 0.27 	& 0.9 	& OMB 		& 3.184 & 5.28 	& 2.20 & 2.22 & 99 	& 2013 & PS 		& Visual & 1 	 & 2013 	& & No 	& [\ref{P2013R3}]\\
P/2015 X6 			& Aeolia		& 2.75	& 0.17 	&  4.6	& MMB 		& 3.318 & 4.57 	& 2.28 & 2.64 & 62 	& 2015 & PS 		& Visual & 1 	 & 2015 	& & No 	& [\ref{P2015X6}]\\
P/2016 G1 			& Adeona		& 2.58	& 0.21	& 11.0	& MMB 		& 3.367 & 4.15 	& 2.04 & 2.52 & 56 	& 2016 & PS 		& Visual & 1 	 & 2016 	& & No 	& [\ref{P2016G1}]\\
P/2016 J1 			& Theobalda	& 3.17 	& 0.23 	& 14.3 	& OMB 		& 3.113 & 5.65 	& 2.45 & 2.46 & 99 	& 2016 & PS 		& Visual & 1 	 & 2016 	& 	& No 	& [\ref{P2016J1}]\\
\enddata
\tablenotetext{}{}
	
	$^\mathrm{\ref{rfirstact}}$Year activity discovered. %
	$^\mathrm{\ref{rfac}}$Facility originally reporting activity. 
	$^\mathrm{\ref{ract}}$Number of times object reported active. 
	$^\mathrm{\ref{rLast}}$As of January 2018 submission. %
	$^\mathrm{\ref{rrefs}}$Object-specific references in Appendix. %
	CSS:Catalina Sky Survey; %
	ESO:European Space Observatory 1-metre Schmidt %
	HTP:Hawaii Trails Project; %
	LINEAR:LIncoln Near-Earth Asteroid pRogram; %
	LSSS:La Sagra Sky Survey; %
	MPCAT:Minor Planet Catalog %
	PS:Pan-STARRS; %
	SS:Siding Spring; %
	SW:Spacewatch %
	
\end{deluxetable*}


\begin{table}
	\caption{AA Mass-loss Mechanisms}
	\centering
	\begin{tabular}{lrr}
	\hline\hline
		Suspected Mechanism & \ N$^*$ & \%\\
		\hline
		Sublimation & 15 & 44\\
		Rotational Breakup & 7 & 21\\
		Impact\ /\ Collision & 4 & 12\\
		Thermal Fracturing & 1 & 3\\
		Cryovolcanism & 1 & 3\\
		Binary Interaction & 1 & 3\\
		Unknown & 5 & 15\\
	\end{tabular}\\
	\raggedright
	$^*$ Objects with multiple mechanism are counted more than once; objects listed in Table \ref{Table:TheAAs} as candidates were not included in this computation.
	\label{Table}
\end{table}

Active asteroids appear to have tails like comets (Figure \ref{ExampleAsteroid}) but follow orbits predominately within the main asteroid belt. Although the first active asteroid (Wilson-Harrington) was discovered in 1949 \citep{Cunningham:1950wt}, 27 of the 31 objects (87\%) were identified as active in the last decade (Table \ref{Table:TheAAs}). Asteroid activity is thought to be caused by several different mechanisms, combinations of which are undoubtedly at work (e.g., an impact event exposing subsurface ice to sublimation). The number of times (i.e., orbits) an object has displayed activity (Table 1: Act.) is especially diagnostic of the mechanism (Table 1: Cause). A singular (non-recurring) event likely originates from an impact event, e.g., (596) Scheila. Rotational breakup, as in P/2013 R3 of Figure \ref{ExampleAsteroid}, may be a one-time catastrophic event, or a potentially repeating event if, for example, only a small piece breaks free but the parent body remains near the spin breakup limit. Ongoing or recurrent activity has been observed $\sim$15 times, e.g., 133P/Elst-Pizarro, and is suggestive of sublimation or, in the case of (3200) Phaethon, thermal fracture. These last two mechanism (sublimation and thermal fracture) should be more likely to occur when an object is closer to the Sun, i.e. perihelion (Table 1:$q$). The Sun-object distance (Table 1: $R$) indicates the absolute distance, but it is can be simpler to consider how close (Table 1: \%$_\mathrm{peri}$) to perihelion the object was when activity was first observed (Table 1: 1$^\mathrm{st}$Act), where 100\% represents perihelion ($q$) and 0\% indicates aphelion:

\begin{equation}
	\mathrm{\%}_\mathrm{peri} = \left[1-\left(\frac{d_\mathrm{disc}-d_\mathrm{peri}}{d_\mathrm{ap}-d_\mathrm{peri}}\right)\right]\cdot 100\mathrm{\%}
\end{equation}

\noindent where $d_\mathrm{disc}$ is the heliocentric object distance at the activity discovery epoch, $d_\mathrm{peri}$ the perihelion distance, and $d_\mathrm{ap}$ the aphelion distance.

While the term ``Main Belt Comets'' often refers to this sublimation-driven subset of active asteroids, we use the more inclusive ``active asteroid'' term throughout this paper. We aimed to include all objects termed ``active asteroids'' in the literature for completeness, but we only include objects which have provided observable signs of activity. Objects known to host surface water ice but which have yet to shown signs of activity, such as (24) Themis \citep{Rivkin:2010ge, Campins:2010fi}, are outside the scope of this paper.

Orbital characteristics also provide insight into the dynamical evolution and even the composition of an object. Objects with conspicuously similar orbital properties may have originated from a catastrophic disruption event that created a family (Table 1:Family) of asteroids \citep{Hirayama:1918jb}. More generally, asteroids can be categorized (Table 1:Orb.) as interior to the Main Asteroid Belt, within the Main Asteroid Belt (and further subdivided into inner, mid, and outer main belt as IMB, MMB, and OMB respectively), or exterior to the Main Asteroid Belt (e.g., Kuiper belt). Objects interior to the Main Asteroid Belt, including Near Earth Objects (NEOs), include Earth-crossing (Apollo), Earth-orbit nearing (Amor), and Mars-crossing asteroids. Objects whose orbits are similar to Comet 2P/Encke are said to be Encke-type.
 
The Tisserand parameter $T_\mathrm{J}$ (Table \ref{Table:TheAAs}$T_J$) describes the degree to which an object's orbit is influenced by Jupiter:

\begin{equation}
	T_\mathrm{J} = \frac{a_\mathrm{J}}{a} + 2\sqrt{\left(1-e^2\right)\frac{a}{a_\mathrm{J}}}\cos(i).
	\label{eq:TJ}
\end{equation}

\noindent The orbital elements are given by $a_\mathrm{J}$ the orbital distance of Jupiter (5.2 AU), plus the semi-major axis $a$, eccentricity $e$, $i$ the inclination (Table 1). For the case where $a=a_\mathrm{J}$ you can see $T_\mathrm{J}=3$. Asteroids in the main-belt are typically inside the orbit of Jupiter (i.e. $a<a_\mathrm{J}$) and usually have $T_\mathrm{J}>3$ \citep{Jewitt:2014vt}; however, as Equation \ref{eq:TJ} indicates, it is the combination of all three free parameters ($a$, $e$, $i$) which describes the magnitude of Jovian influence on the object's orbit. One active asteroid definition also constrains membership to objects whose orbits are interior to Jupiter but whose Tisserand parameters are $>$ 3.08 \citep{Jewitt:2014vt}.

Objects not identified in the literature as active asteroids, yet still appear orbitally asteroidal (e.g., Comet 2P/Enke), are not included in this paper, but objects with $T_\mathrm{J}<3$ which are identified in the literature as active asteroids (e.g., (3552) Don Quixote), are included; see e.g., \cite{Hsieh:2006dk,Tancredi:2014kl} for further discussion on distinguishing objects within this regime.


We would like to understand active asteroids in part because they may hold clues about solar system formation and the origin of water delivered to the terrestrial planets. The recent discovery of interstellar asteroid \omuamua{} \citep{Bacci:2017td} intensifies interest in understanding our own indigenous asteroid population in order to better understand and characterize ejectoids we encounter in the future, an estimated decadal occurrence \citep{Trilling:2017ia}. There has also long been an interest mining asteroids for their metals, and water could prove an invaluable resource providing, for example: energy, rocket fuel, breathable oxygen, and sustenance for plant and animal life \citep{OLeary:1977eq,Dickson:1978dm,Kargel:1994in,Forgan:2011el,Hasnain:2012jo,Lewicki:2013gn,Andrews:2015jq}.

Our knowledge of active asteroids has been limited due to small sample size: only $\sim$20 active asteroids have been discovered to date \citep{Jewitt:2015fl}. As such, the statistics presented in Table \ref{Table} are poorly constrained (e.g., the thermal fracturing rate is based upon a single object: (3200) Phaethon). Spacecraft visits have been carried out or planned to a number of the active asteroids (Table 1: Visit), and while we may learn a great deal from these individual objects, spacecraft visits will not substantially increase the number of known active asteroids. While spectroscopy has recently shown potential for discovering activity, the overwhelming majority of activity detections have been made by visual examination (Table 1:Method). One notable exception was the 1984 (2201) Oljato outburst first detected by magnetic field disturbances (2201) Oljato outburst \citep{Russell:1984eq}.

\begin{table}
	\small
	\centering
	\caption{Surveys which have discovered AAs.}
	\begin{tabular}{lccc}
		\toprule
		AA Discovered by & AAs & Limit & Operation\\
		\hspace{4mm}(survey name) & (N) & (mag) & (years)\\
		\hline
		Catalina Sky Survey & 5 & 22$^a$ & 1998$^f$ --\\
		La Sagra Survey & 2 & 17$^b$ & 2008$^g$--\\
		LINEAR & 1 & 19.6$^c$ & 1997$^h$ --\\
		Pan-STARRS & 8 & 22.7$^d$ & 2008$^i$ --\\
		Spacewatch & 2 & 21.7$^e$ & 1981$^j$ --\\
		Total & 18 & $\cdots$ & 98\\
		\hline
	\end{tabular}
	
	\raggedright
	$^a$\cite{2009ApJ...696..870D}; 
	$^b$estimated from aperture; 
	$^c$\cite{2011AJ....142..190S,Stokes:2000fc}; 
	$^d$\cite{Chambers:2016vk}; 
	$^e$\cite{2007AJ....133.1247L}; 
	$^f$\cite{1998BAAS...30.1037L}; 
	$^g$\cite{Stoss:2011vl}; 
	$^h$\cite{Stokes:2000fc}; 
	$^i$\cite{2008LPICo1405.8301J}; 
	$^j$\cite{Gehrels:1981kf}
	
	\label{Surveys}
\end{table}

\begin{table*}
	\centering
	\caption{Active Asteroid Hunting Surveys \& Occurrence Rate Estimates}
	\small
	\begin{tabular}{p{4.2cm}p{2.2cm}p{1.2cm}p{1.79cm}cp{0.9cm}rl}
		\toprule
		Survey & Source & Zone & Activity \footnotesize{($N$ per $10^6$)} & $N^\dag$ & Limit (mag) & Objects & Method\\
		\hline
			\cite{Cikota:2014ds} & MPC & MBA & \hspace{6mm}$\cdots$ & 1 & 16.7 & 330K & Photometric Excess\\
			\cite{2010Icar..210..998G} & CFHT & MBA & \hspace{6mm}$40\pm 18$ & 3 & 22.5$^a$ & 25K & By-Eye\\
			\cite{Hsieh:2009gy} & HTP & OMB & \hspace{6mm}$\cdots$ & 1 & 26 & 600 & By-Eye\\
			\cite{Hsieh:2015jo} & \footnotesize{Pan-STARRS} & OMB & \hspace{6mm}$96$ & 4 & 22.6 & 300K & PSF\\
			SAFARI (this work) & DECam & MBA & \hspace{6mm}$80$ & 1 & 24.3 & 11K & By-Eye\\
			\cite{Sonnett:2011ip} & TALCS & MBA & \hspace{2mm}$<2500$ & 0 & 24.3 & 1K & Excess Sky Flux\\
			\cite{Waszczak:2013be} & PTF & MBA & \hspace{2mm}$<30$ & 0 & 20.5 & 220K & Extendedness\\
		\hline
	\end{tabular}
	\raggedright
	CFTS: Canada-France-Hawaii Telescope; DECam: Dark Energy Camera; HTP: Hawaii Trails Project; MPC: Minor Planet Center; PTF: Palomar Transient Factory; Pan-STARRS: Panoramic-Survey Telescope And Rapid Response System; TALCS: Thousand Asteroid Light Curve Survey; MBA: Main Belt Asteroids; OMB: Outer Main Belt; $^\dag$Includes known AAs;
	$^a$\cite{2009Icar..201..714G}; PSF: Point Spread Function
	\label{aahunts}
	\label{ActivityRates}
\end{table*}

We chose to visually examine (``by-eye'') images of active asteroids because this technique has so far produced the greatest yield. Other methods have been applied (Table \ref{Surveys}) but with varied degrees of success. \cite{Cikota:2014ds} examined a large number of objects and searched for unexpected deviations in object brightness; this technique positively identified one known active asteroid, but (so far) the other candidates (\ref{Table:TheAAs}) have not been observed to be active. \cite{Sonnett:2011ip} examined the regions immediately surrounding asteroids, searching for photometric excess (i.e., a photon count above the sky background level). \cite{Waszczak:2013be} formulated a way to quantify ``extendedness'' of  Palomar Transient Factory objects, with a 66\% comet detection rate and a 100\% Main Belt Comet detection efficiency. \cite{Hsieh:2015jo} compared point spread function (PSF) widths between background stars and other objects and flagged exceptionally large PSF radii for further follow-up. All of the aforementioned techniques rely upon visual inspection for confirmation of activity. Spectroscopic detection of activity has also been carried out (Table \ref{Table:TheAAs2}), but so far only (1) Ceres has been observed to be visually active in follow-up, and, in that case, in situ by the Rosetta spacecraft orbiting it. Hayabusa 2 recently arrived at (162173) Ryugu but as of yet no tail or coma has been observed.

Conservative activity occurrence rates of $>$1 in 10,000 are constrained by the magnitude limits of prior surveys \citep{Jewitt:2015fl}. We reached past the 17-22.7 magnitude limits of previous large-sky surveys (Table \ref{Surveys}) by making use of existing Dark Energy Camera (DECam) data \citep{Sheppard:2016jf} probing a magnitude fainter than other large-sky active asteroid survey. Note that while we are sensitive to more distant populations (e.g., Centaurs, Trans-Neptunian Objects), 99.7\% of our population is from the main asteroid belt.

We set out to determine the viability of DECam data for locating active asteroids. We aimed to create a novel, streamlined pipeline for locating known asteroids within our dataset. We planned to examine our new library of asteroid thumbnails to find active asteroids and to test published asteroid activity occurrence rates (Table \ref{ActivityRates}). We applied our technique to \fitscount{} DECam images ($\sim$5 Tb) to produce \allthumbs{} thumbnail images comprising \uniquethumbs{} unique objects. We examined the asteroid thumbnails by-eye to identify signs of activity. We show our technique can be applied to an orders-of-magnitude larger publicly-available dataset to elevate active asteroids to a regime where they can be studied as a population.

\section{Methods}
\label{methods}

\begin{figure}
  \centering
	\includegraphics[width=0.85\linewidth]{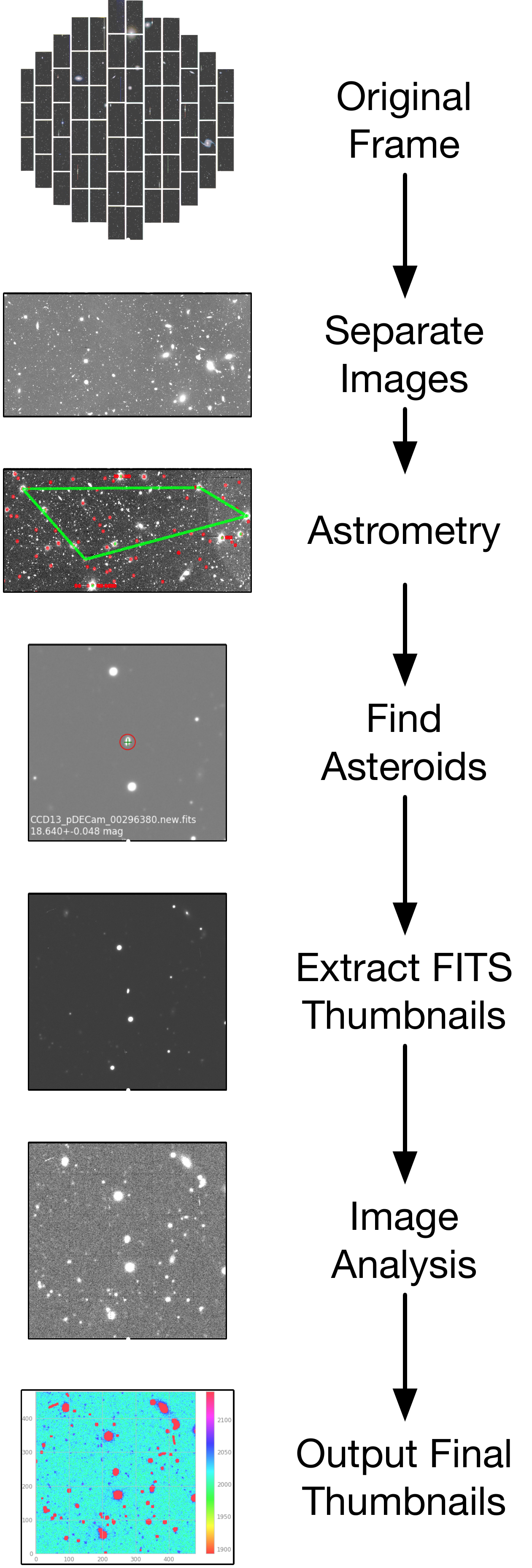}
  \caption{The SAFARI workflow.}
  \label{flow}
\end{figure}

\subsection{Dark Energy Camera} We made use of data taken by the Dark Energy Camera (DECam) instrument on the 4-meter Blanco telescope at the Cerro Tololo Inter-American Observatory in Chile. The instrument has a $\sim$3 square degree field of view, capturing data via a mosaic of 62 charge-coupled device (CCD) chips, each $2048\times 4096$ with a pixel scale of 0.263 arcseconds per pixel \citep{DarkEnergySurveyCollaboration:2016ey}. Our data consisted of $594\times 2.2$ Gb frames in the VR filter ($500\pm10$ nm to $760\pm10$ nm), each containing $62\times 33$ Mb subsets of data, one per CCD. The mean seeing across all images was $1.14\pm 0.13$ arcseconds. We made use of software which required each multi-extension Flexible Image Transport System (FITS) file be split into its 62 constituent parts, which we refer to as images for the remainder of this paper. Note: some files contained only 61 chips due to an instrument hardware malfunction.

\subsection{High Performance Computing}

We utilized \textit{Monsoon}, the Northern Arizona University (NAU) High Performance Computing (HPC) computing cluster. \textit{Monsoon} uses the \textit{Slurm Workload Manager} \citep{YooAB:2003ta} software suite to manage the 884 Intel Xeon processors to deliver up to 12 teraflops of computing power. The majority of our tasks each utilized 8 cores and 48 Gb of memory. The online supplement contains the complete listing of requirements necessary for each task.

\subsection{photometrypipeline} We utilized the \textit{photometrypipeline} \citep{Mommert:2017jy} software package to carry out source extraction via \textit{Source Extractor} \citep{Bertin:1996hfa,Bertin:2010wt}, photometry and astrometry via \textit{SCAMP} \citep{Bertin:2006vk,Bertin:2010wt}, and asteroid identification via \textit{SkyBot} \citep{Berthier:2006tn} and \textit{Horizons} \citep{Giorgini:2015vs}. We chose the \textit{Anaconda}\footnote{\url{www.anaconda.com}} \textit{Python} programming language distributions (versions 2.7 and 3.5) and the \textit{Python} package \textit{AstroPy} \citep{AstropyCollaboration:2013cd}.

\subsection{Procedure}
\label{procedure}

\begin{figure}
	\centering
	\begin{tabular}{cc}
		\includegraphics[width=0.46\linewidth]{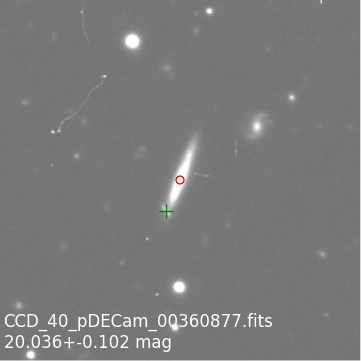} & \includegraphics[width=0.46\linewidth]{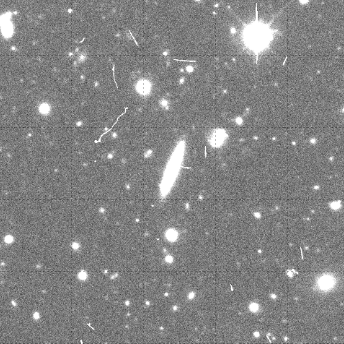}\vspace{-2mm}\\
		\vspace{1mm}
		(\textbf{a}) & (\textbf{b})\\
		\includegraphics[width=0.46\linewidth]{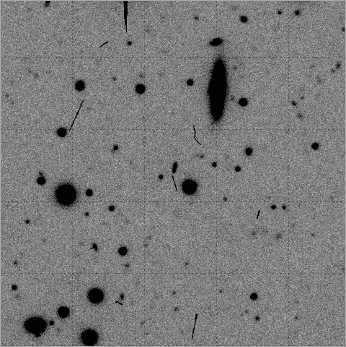} & \includegraphics[width=0.46\linewidth]{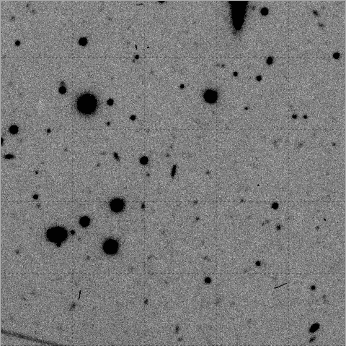}\vspace{-2mm}\\
		(\textbf{c}) & (\textbf{d})\\
	\end{tabular}
	\caption{Two asteroid thumbnail contrast selection approaches are shown in \textbf{a} and \textbf{b}. \textbf{a}) The \textit{photometrypipeline} thumbnail shows increased dynamic range. \textbf{b}) Iterative Rejection sacrifices some dynamic range (notice especially the edges of the center galaxy and the spiral galaxy to its upper-right) in favor of recovering more objects, many of which are not visible in \textbf{a} that can be easily seen in \textbf{b}. \textbf{c} \& \textbf{d}) Asteroid 2012 YU2 is shown in two frames comprising one animated GIF file.}
	\label{AnimatedGif}
	\label{ContrastComparison}
	\label{ExposureTimes}
\end{figure}

\begin{enumerate}[itemsep=2pt,leftmargin=10pt]
	\item \textit{Image Reduction}-- We employed standard image reduction techniques where each frame was bias subtracted, then flat-fielded using a combination of twilight flats and a master flat; full details of our imaging techniques can be found in \cite{Sheppard:2016jf}.
	\item \textit{Splitting Multi-Extension FITS Files}-- DECam produces multi-extension FITS files, where each extension contains data from one CCD; because \textit{photometrypipeline} was incompatible with this format, we split each file into 62 separate FITS files via the \textit{FTOOLS} \citep{Blackburn:1995wn} software package. We replicated global and extension headers for each output file to preserve metadata required for our image processing.
	\item \textit{Coordinate Correction}-- Each DECam image came pre-encoded with right ascension (RA) and declination (Dec) information indicating the coordinates of the telescope pointing center. We shifted the RA \& Dec of each remaining CCD to their true coordinate values. The RA \& Dec offsets used for each CCD are provided with the online supplement.
	\item \textit{World Coordinate System Purging}-- We discovered World Coordinate System (WCS) headers encoded in the FITS files were preventing \textit{photometrypipeline} and/or \textit{astrometry.net} from performing astrometry. We were able to resolve the issue by purging all WCS header information as part of our optimization process. The header record names are listed in the online supplement.
	\item \textit{WCS Population via astrometry.net}-- We installed the \textit{astrometry.net} \citep{Lang:2010cq} v0.72 software suite on \textit{Monsoon}. We processed all \fitscount{} FITS files to retrieve coordinate information for each image by matching the image to one or more index files (catalogs of stars for specific regions of sky, designed for astrometric solving).
	\item \textit{photometrypipeline Image Processing}-- We performed source extraction, photometry, astrometry and image correction via the \textit{photometrypipeline} software suite.
	\item \textit{Identifying Known Asteroids} We identified known asteroid in our data by making use of \textit{pp\_distill}, a module of \textit{photometrypipeline}.
	\item \textit{FITS Thumbnail Generation}-- We extracted the RA, Dec, and $(x,y)$ pixel coordinates of each object. We then produced $480\times480$ pixel, lossless, FITS format asteroid thumbnails, each a small image centered on an asteroid. For cases where the object was too close ($<$240 pixels) to one or more image edges, we found it best to use the \textit{NumPy}\footnote{\url{www.numpy.org}} \textit{Python} routine to ``roll'' the image array; the technique shifts an array as if it were wrapped around a cylinder. For example: array [0, 1, 2, 3] rolled left by 1 would result in array [1, 2, 3, 0].
	\item \textit{Create PNG Thumbnails}-- We used an iterative-rejection technique to compute contrast parameters, then produced Portable Network Graphics (PNG) image files via \textit{MatPlotLib}\footnote{\url{www.matplotlib.org}}.
	\item \textit{Animated GIF Creation} We combined thumbnails of asteroids observed more than once (Figure \ref{DataStats}c) to create animated Graphic Interchange Format (GIF) files (Figure \ref{AnimatedGif}) using the \textit{Python Image Library}\footnote{\url{www.pythonware.com/products/pil/}} software package. There are a number of advantages to this inspection approach, including 1) the opportunity to inspect one asteroid at multiple epochs, 2) activity may not occur at every epoch, and 3) activity may be easier to spot if the inspector has the opportunity to become familiar with an object (e.g., the general shape or streak pattern), even if only briefly.
	\item \textit{Examination of Image Products} -- Three authors served as asteroid thumbnail inspectors. Each inspector conducted a procedure consisting of rapid by-eye examination of asteroid thumbnails and animated GIFs, covering each thumbnail at least once. We flagged thumbnails and animations containing potential active asteroids for a later en masse review.
\end{enumerate}

\section{Results}
\label{results}

\begin{figure*}
	\centering
	\begin{tabular}{cc}
		\begin{tabular}{cc}
			\includegraphics[width=0.2\linewidth]{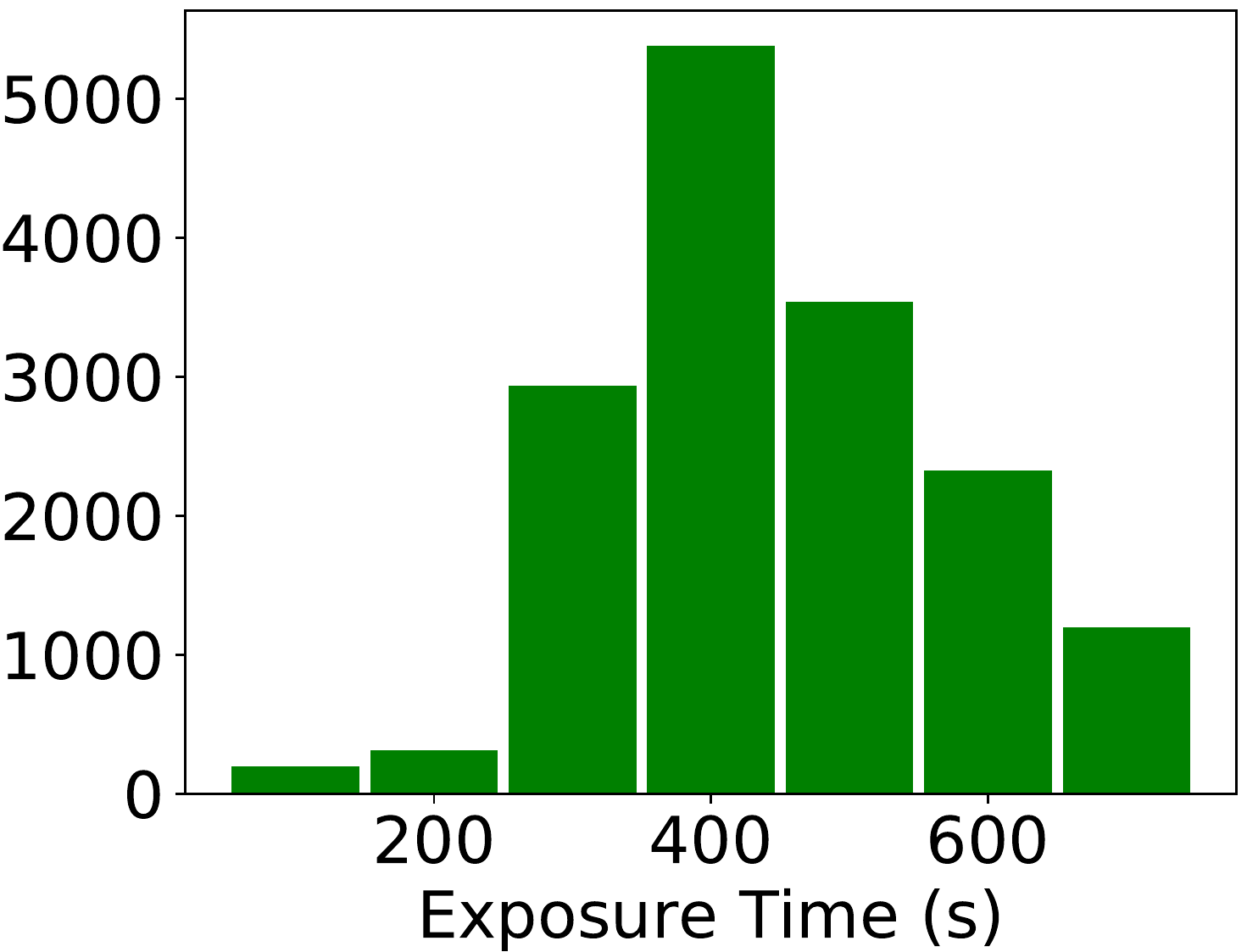} & \includegraphics[width=0.2\linewidth]{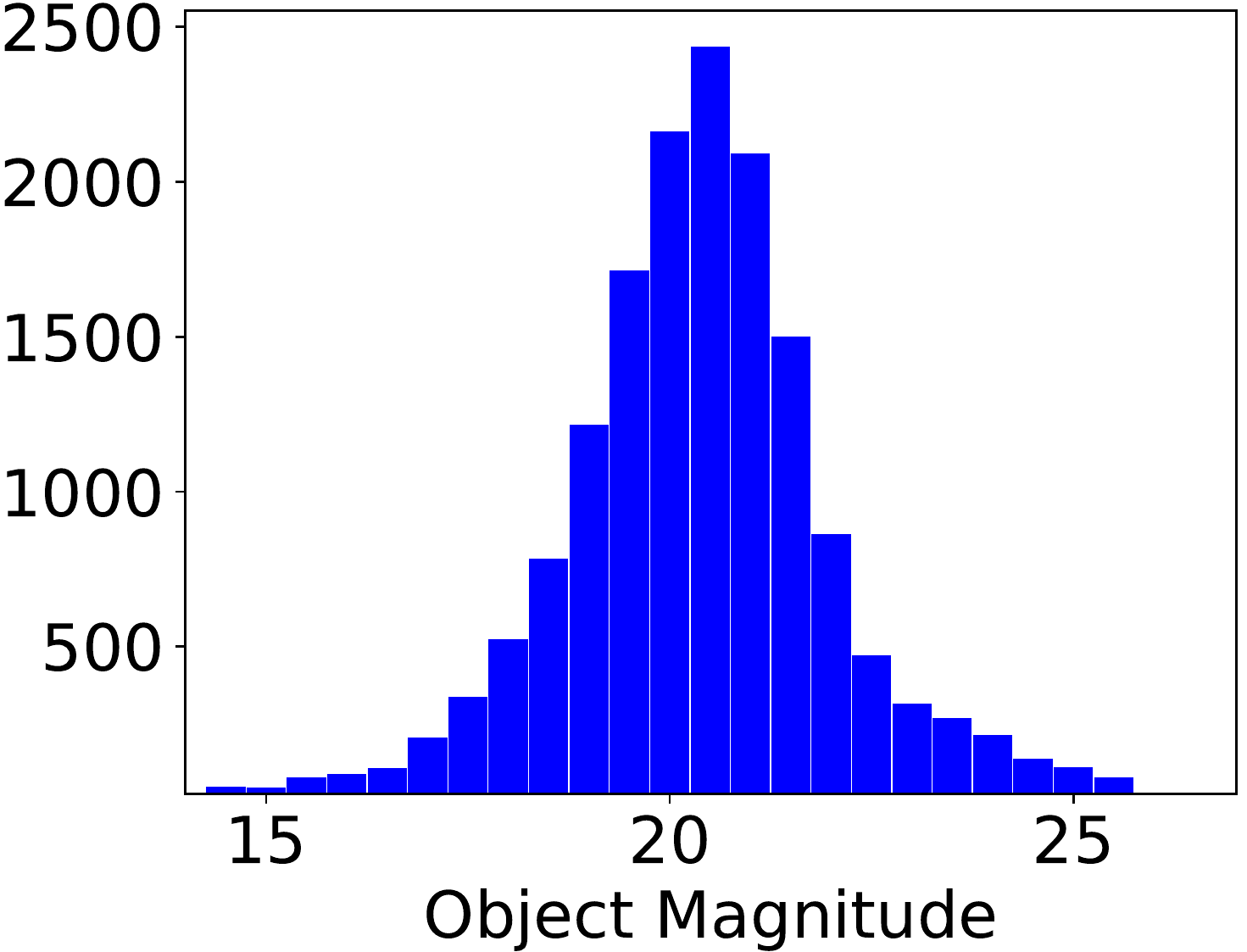}\vspace{-2mm}\\
			(\textbf{a}) & (\textbf{b})\vspace{1mm}\\
			\includegraphics[width=0.19\linewidth]{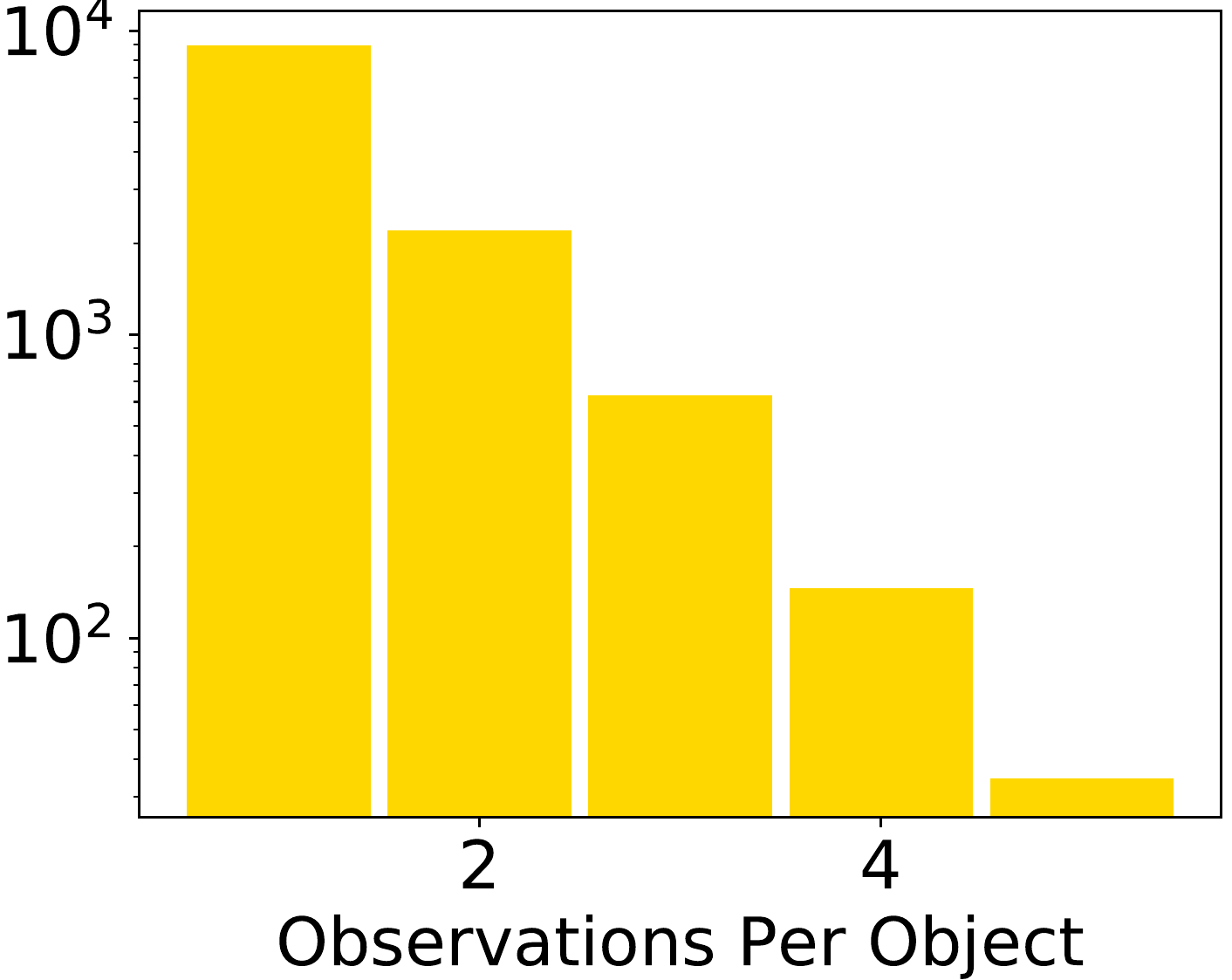} & \includegraphics[width=0.21\linewidth]{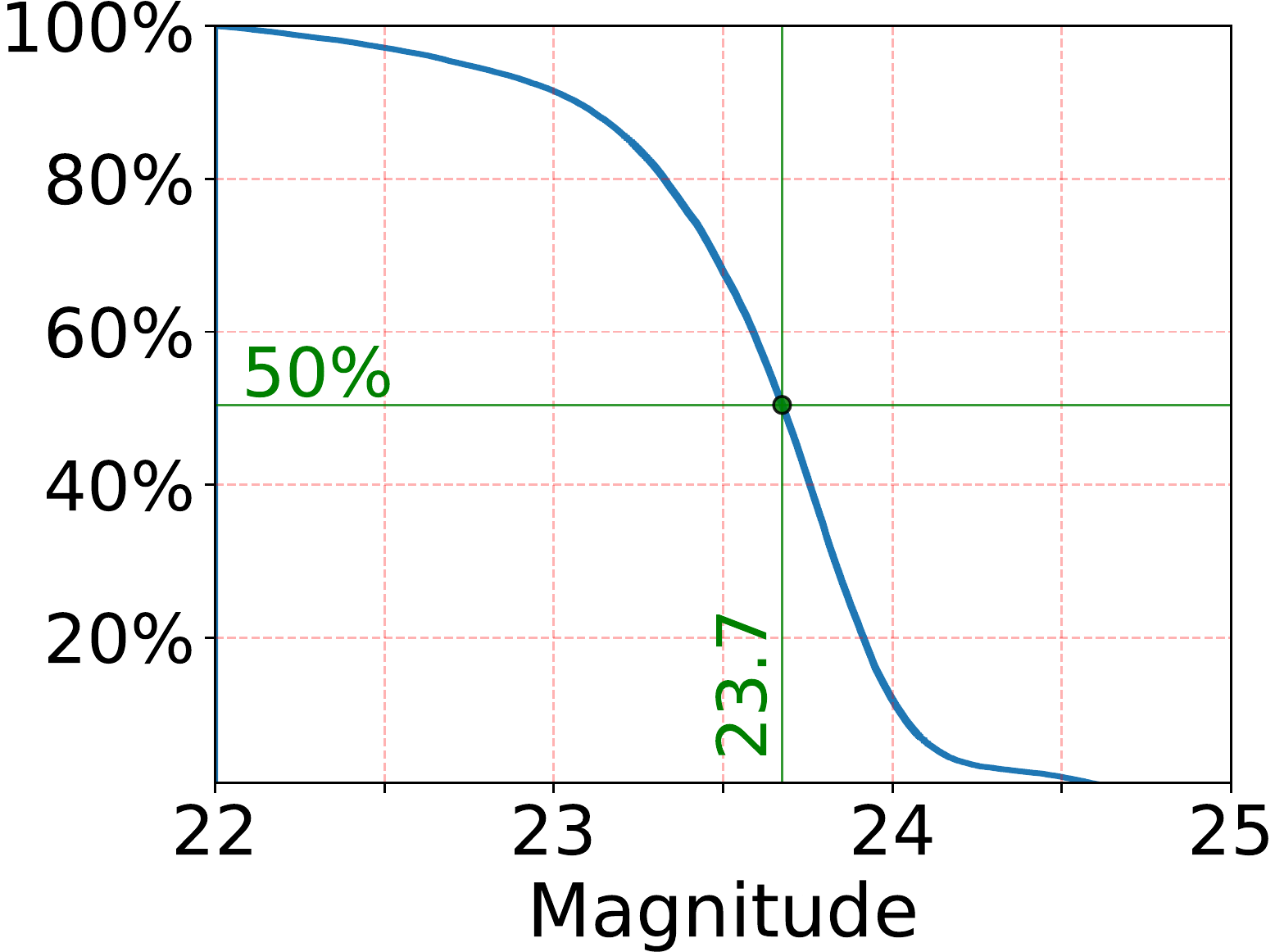}\vspace{-2mm}\\
			(\textbf{c}) & (\textbf{d})\\
		\end{tabular} & \hspace{-10mm}
		\begin{tabular}{c}
			\includegraphics[width=0.5\linewidth]{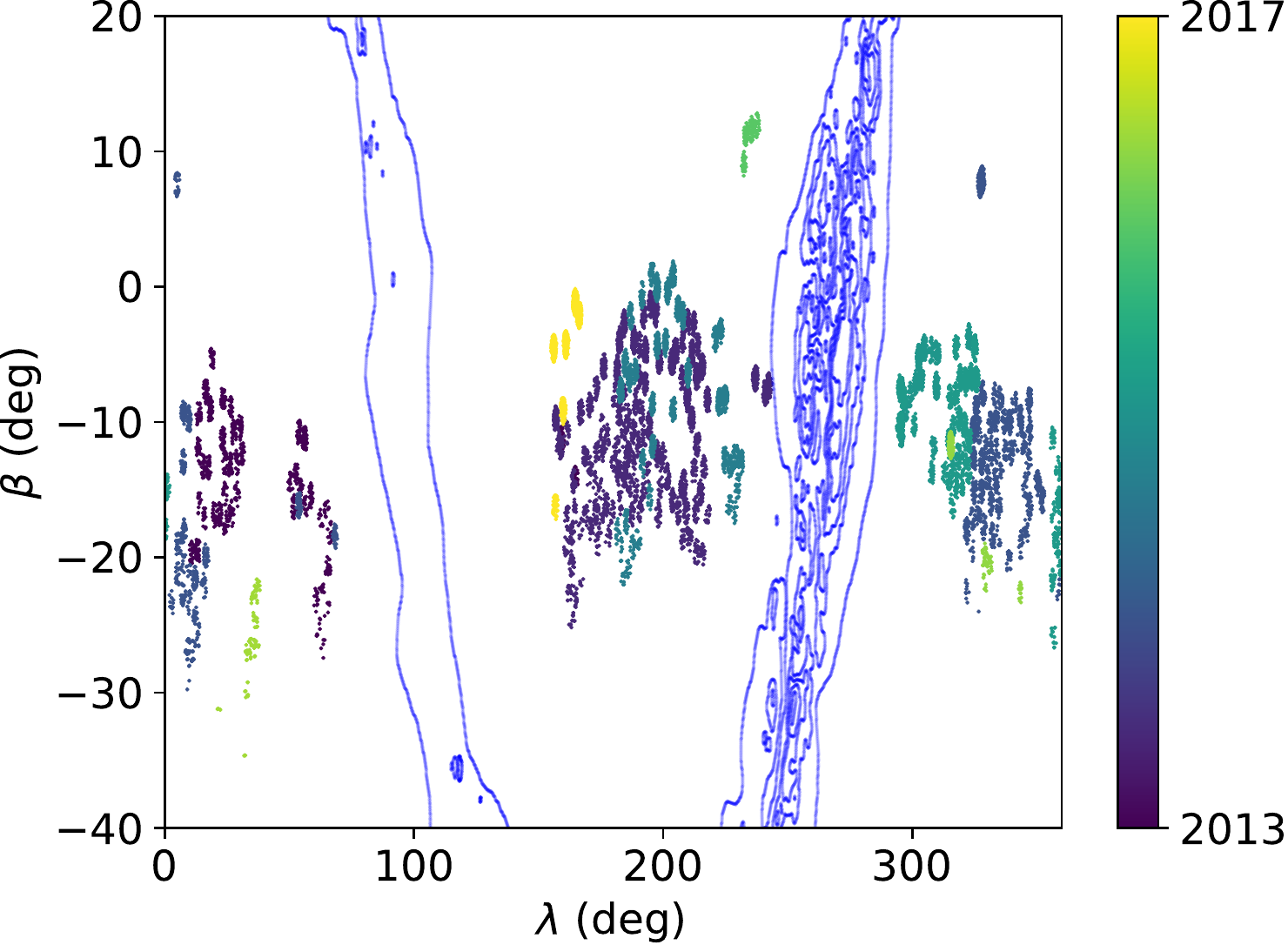}\vspace{-2mm}\\
			(\textbf{e})\\
		\end{tabular}
	\end{tabular}
	\caption{\textbf{a}) Exposure time distribution in our data. \textbf{b}) Histogram of apparent magnitudes for known asteroids we identified in our dataset. \textbf{c}) Observations per object; the \allthumbs{} asteroid thumbnails contained \uniquethumbs{} unique objects, \gifcount{} of which were observed more than once. \textbf{d}) Cumulative histogram showing the depth of magnitudes (stars and asteroids) found in our dataset. 50\% of our images reached a magnitude of $m_\mathrm{R}=23.7$. Sources with a signal-to-noise ratio of $<$5:1 were not included. \textbf{e}) Asteroids encountered shown in geocentric ecliptic space, where $\lambda$ and $\beta$ are the ecliptic longitude and latitude, respectively. Distinct patches sum to $\sim$1000 $\mathrm{deg}^2$, as described in the text. Milky Way coordinates were retrieved from the \textit{D3-Celestial} (\url{http://ofrohn.github.io}) software suite.}
	\label{DataStats}
\end{figure*}

\textit{Pipeline} -- We created a pipeline (Figure \ref{flow}) that takes as its input DECam multi-extension FITS files, and returns individual asteroid thumbnails and animated GIF files. The initial total compute time requested across all tasks was 13,000 hours (1.5 compute-years), but after optimization (see Optimization section below) only $\sim$500 compute hours were required. See the online supplement for a comprehensive table of resources utilized during this project.

\textit{Image Products} We extracted \allthumbs{} asteroid thumbnails from \fitscount{} DECam images ($\sim$2 Tb total). Most of our data consisted of exposure times $>$300s (Figure \ref{DataStats}a). These longer integration times allowed us to probe deeper (fainter), with asteroids captured down to $25^\mathrm{th}$ magnitude (Figure \ref{DataStats}b). Each of the \uniquethumbs{} unique objects identified in our data were observed between 1 and 5 times, with \gifcount{} objects imaged more than once (Figure \ref{DataStats}c). 

To compute our coverage area on sky (depicted in Figure \ref{DataStats}e) we employed a nearest neighbor algorithm to identify the distinct (non-overlapping) regions of our dataset. Two fields were considered overlapping if their center-to-center distance was $<1.8$ degrees, the width of one DECam field. We computed our coverage to be $\sim200$ distinct 3 $\mathrm{deg}^2$ patches comprising $\sim$1000 square degrees.

\begin{figure}
	\centering
	\includegraphics[width=1.0\linewidth]{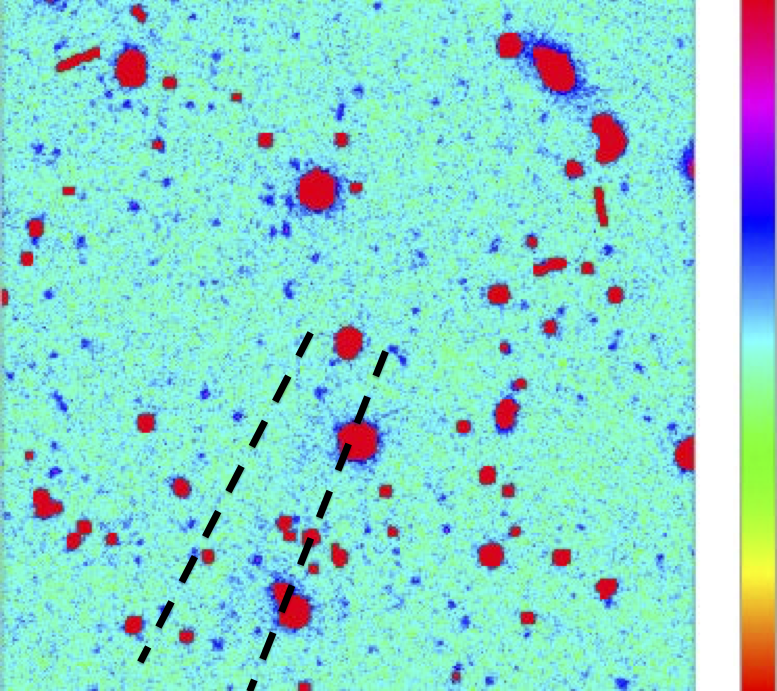}
	\caption{Asteroid (62412) shown with the ``hsv'' colormap and Mitchell interpolation. The asteroid is at the center of the frame and the tail can be seen between the dashed lines.}
	\label{fig:62412}
\end{figure}

\textit{Active Asteroids} -- We imaged one asteroid previously discovered to be active \citep{Sheppard:2015cw}: (62412). The object shows activity in our image (Figure \ref{fig:62412}; see the online supplement for additional image color map and interpolation permutations) and we were able to identify activity in two other DECam frames that were not part of this work. \cite{Sheppard:2015cw} confirmed activity with Magellan Telescope follow-up observations. We also imaged two other objects listed as active: (1) Ceres and (779) Nina but neither showed signs of activity.

\textit{Optimization} --  The final pipeline resulted from a series of iterative optimizations carried out with a subset of our large dataset. These optimizations produced order-of-magnitude reductions in compute time, and improved successful pipeline completion from the initial $\sim$35\% to the final 94\%. The implemented optimizations and their results are broken down below by number (matched to the corresponding procedure number of Section \ref{procedure}). The final optimized \textit{Slurm} parameters used on \textit{Monsoon} can be found in the online supplement.

\begin{enumerate}[itemsep=2pt,leftmargin=10pt]
	\item \textit{Image Reduction}: No optimization needed.
	\item \textit{File-Splitting}: Splitting each multi-extension FITS file into 62 separate FITS files resulted in a larger number of smaller tasks which were better suited for parallel processing.
	\item \textit{Coordinate Correction}: Coordinate corrections proved cumbersome and inefficient, so we added \textit{astrometry.net} to our pipeline.
	\item \textit{WCS Purging}: We identified mismatched distortion coefficients as the primary culprit behind roughly 1/3 of our images failing \textit{photometrypipeline} analysis. We purged all World Coordinate System (WCS) headers, allowing us to employ \textit{astrometry.net} which increased our overall throughput and output.
	\item \textit{astrometry.net Astrometry}: We cached all ($\sim$32 Gb) astrometry index files (described in Section \ref{procedure} item 5) locally so \textit{astrometry.net} would not be dependent on the speed of the internet connection and file host. We optimized the \textit{astrometry.net} computation by supplying the following parameters we extracted from our FITS files. Providing a pixel scale range ($\sim$0.25'' to $\sim$0.28'') and RA/Dec values narrowed the range of indices that required searching. We found a 15'' search radius further reduced computation time without impacting image recognition efficacy. We disabled \textit{astrometry.net} plotting due to a \textit{Slurm} incompatibility, and computation time decreased further still. We found submitting \textit{astrometry.net} ``solve-field'' tasks directly to \textit{Slurm} was much faster. All but 41 images successfully matched for astrometry on first pass, and we improved \textit{astrometry.net} image recognition speed roughly tenfold.
 	\item \textit{photometrypipeline}: Proper configuration of prerequisite software and \textit{photometrypipeline} proved crucial; the online supplement contains the necessary parameters we used. We made minor modifications to the \textit{photometrypipeline} code, described in the online supplement. We found out \textit{astropy} was using home directory temporary storage space, a fatal error for systems with enforced quotas; the home storage space was also slower than the scratch space. Proper configuration reduced computation time and increased the pipeline success rate.
 	\item \textit{Known Asteroid Identification} We added an initial \textit{SkyBot} query to identify the asteroids within each image. We then populated the requisite \texttt{OBJECT} FITS header keyword in each of our images, thereby enabling us to call \textit{Horizons} to locate asteroids in our images and provide accurate astrometry. Prepending the \textit{SkyBot} query and populating the \texttt{OBJECT} keyword enabled us to run asteroid identification tasks in parallel, reducing processing time by three orders-of-magnitude.
	\item \textit{FITS Thumbnails}: We ``rolled'' images (described in Section \ref{procedure} item 8) so we could create full-sized ($480\times480$ pixel) thumbnails. While thumbnails sometimes looked peculiar when rolled, this method preserved image statistics used to compute the narrow range of contrast achieved in the next section.
	\item \textit{PNG Thumbnails}: While \textit{photometrypipeline} does output thumbnails by default, we were unable to see enough detail with the default scaling. Therefore, we employed an iterative rejection technique. Figures \ref{ContrastComparison} a and \ref{ContrastComparison} b compare the two contrast ranges. For each of the \allthumbs{} asteroid thumbnails, we chose to output different colormap/interpolation combinations: two modes of interpolation (Mitchell-Netravali balanced cubic spline filter and one set unfiltered), each in 11 color schemes (afmhot, binary, bone, gist\_stern, gist\_yarg, gray, hot, hsv, inferno, Purples, and viridis), examples of which are shown in the online supplement. 
The optimized dynamic ranges allowed faint trails to become more visible. These colormap/interpolation schemes gave us, as thumbnail inspectors, the ability to choose a comfortable theme for use while searching thumbnails for asteroid activity, thereby increasing our productivity.
	\item \textit{Animated GIFs}: We produced animated GIFs enabling an alternative inspection format.
	\item \textit{Examination}: We uncovered common sources of false positives (discussed in Section \ref{FalsePositives}) and incorporated their presence into our visual examination procedures, resulting in a streamlined examination process while simultaneously reduced the number of false-positives.
\end{enumerate}

\section{Discussion}
\label{discussion}

We set out to determine if DECam data would provide a suitable pool from which to search for active asteroids. We crafted a method to extract asteroid thumbnails from DECam data, and the large number of asteroids encountered (\uniquethumbs{}) along with the exceptional depth our images probed (Figures \ref{DataStats}b and \ref{DataStats}d) indicate our data are well-suited to locating active asteroids. 

\subsection{Population Traits}

\begin{table}
	\centering
	\small
	\caption{SAFARI Asteroid Populations}
	\begin{tabular}{lclccrr}
		\hline\hline
		Zone 		& $R_i$ & $a_{p_i}$ & $a_{p_o}$ & $R_o$ & \multicolumn2c{SAFARI}\\
		      		& \footnotesize{(A:J)} & \footnotesize{(au)}    & \footnotesize{(au)} & \footnotesize{(A:J)} & (N) & (\%)\ \\
		\hline
		Int.  		& $\cdots$ 	& 0     & 2.064    & 4:3 	  &	115   &  1\\
		IMB   		& 4:3 		& 2.064 & 2.501    & 3:1 	  & 3,605 & 26\\
		MMB   		& 3:1 		& 2.501 & 2.824    & 5:2 	  & 5,358 & 39\\
		OMB   		& 5:2 		& 2.824 & 3.277    & 2:1 	  & 4,599 & 33\\
		Ext.		& 2:1 		& 3.277 & $\infty$ & $\cdots$ & 162   &  1\\
		Total$^*$	& $\cdots$	& $\cdots$ & $\cdots$ & $\cdots$ & 13,839 & 100\\
		\hline
	\end{tabular}
	\label{tab:zones}
	
	\raggedright
	\footnotesize{Int., Ext.,: Interior, Exterior to the main belt\\
	IMB, MMB, OMB: Inner, Mid, Outer Main Belt;\\	
	$a_{p_i}$,$a_{p_o}$: inner, outer proper semi-major axis;\\
	A:J Asteroid:Jupiter; $R_i$, $R_o$: inner/outer resonances\\
	$^*$Not included: 791 objects with unknown parameters.}
\end{table}

As indicated by Figures \ref{DataStats}a-d, the population imaged during our survey were subject to selection effects caused by the depth ($\bar{m}_\mathrm{R}=23.7$) of our survey (e.g., closer objects would have appeared as long trails which would have been difficult to identify with our pipeline). We classified the objects following the procedure of \cite{Hsieh:2018kk}; we categorized our population as Inner Main Belt (IMB), Mid Main Belt (MMB), and Outer Main Belt (OMB), plus two additional regions: ``Interior'' (to the IMB) and ``Exterior'' (to the OMB). Table \ref{tab:zones} indicates the boundaries, along with their Asteroid:Jupiter (A:J) resonances.

The synthetic proper semi-major axis $a_p$ aims to minimize the influence of transient perturbations \citep{Knezevic:2000uf}. We made use the \textit{AstDyn-2}\footnote{\url{http://hamilton.dm.unipi.it/astdys}} online catalog service \citep{Knezevic:2003hw} in determining proper orbital parameters for asteroids in our dataset (Table \ref{tab:zones}).



Our target (object) aperture photometry was computed with a fixed diameter of 10 pixels, though photometric calibration was performed with an aperture radius determined by curve-of-growth analysis (see \citet{Mommert:2017jy} for details). To determine the surface brightness limit of our catalog we first computed the limit $SB$ of each image

\begin{equation}
	SB_\mathrm{lim} = \frac{\sum_{k=1}^{k=N} \left(m_{0_k} - 2.5 \log_{10}\left(n \sigma_{\mathrm{bg}_k}\sqrt{1/A}\right)\right)}{N},
\end{equation}

\noindent where $m_0$ is the photometric zero point (determined by \textit{PhotometryPipeline}), $n$ the order of detection level for background noise standard deviation $\sigma_\mathrm{bg}$, and $A$ is the area of one pixel in square arcseconds \citep{Hsieh:2018up}. The DECam camera had a pixel scale of 0.263''/pixel, give a pixel area

\begin{equation}
	A = (0.263'')^2 = 0.069169\ \mathrm{arcseconds}^2.
\end{equation}


For our surface brightness analysis we made use of $N=32,790$ chips for which we had been able to determine a photometric zero point. We computed the $3\sigma$ mean surface brightness limit of our dataset to be ${SB}_\mathrm{lim}=27.9\pm 1.2$ mag/$\mathrm{arcsec}^2$.

\subsection{Occurrence Rates}
We also aimed to validate the published asteroid activity occurrence rates of Table \ref{ActivityRates}. Occurrence rates have been conservatively set at 1 in 10,000 (for all main belt asteroids), with the limiting magnitude of surveys the primary bottleneck. As shown in Figure \ref{DataStats}d, the DECam instrument reaches an average magnitude of 24 \citep{Sheppard:2016jf}, an unprecedented depth for large area active asteroid surveys. While our complete dataset was consistent with the 1:10,000 activity occurrence rate estimate, it is somewhat surprising we did not discover additional asteroidal activity. 

\cite{Hsieh:2015jo} postulated many active asteroids could be continuously active throughout their orbits (not just at perihelion), but with weaker activity. We expected then to find active asteroids more frequently in our search, given the objects we observed were indeed of a fainter magnitude (Figure \ref{DataStats}b), though our outer main belt occurrence rate ($\sim$1:4000) was slightly higher than that reported by \cite{Hsieh:2015jo} which is in line with their prediction. Small number statistics may have contributed to the possible discrepancy, and it is plausible we missed activity indications due to the limitations of visual inspection which were further compounded by an increased prevalence of background sources compared to shallower surveys. The use of a point spread function (PSF) comparison technique (e.g., \citet{Hsieh:2015jo} or a photometric search (e.g., \citet{Cikota:2014ds}) could help us identify candidates, features we plan to investigate in future work.

\subsection{False Positives}
\label{FalsePositives}

\begin{figure}
	\centering
			\includegraphics[width=1.0\columnwidth]{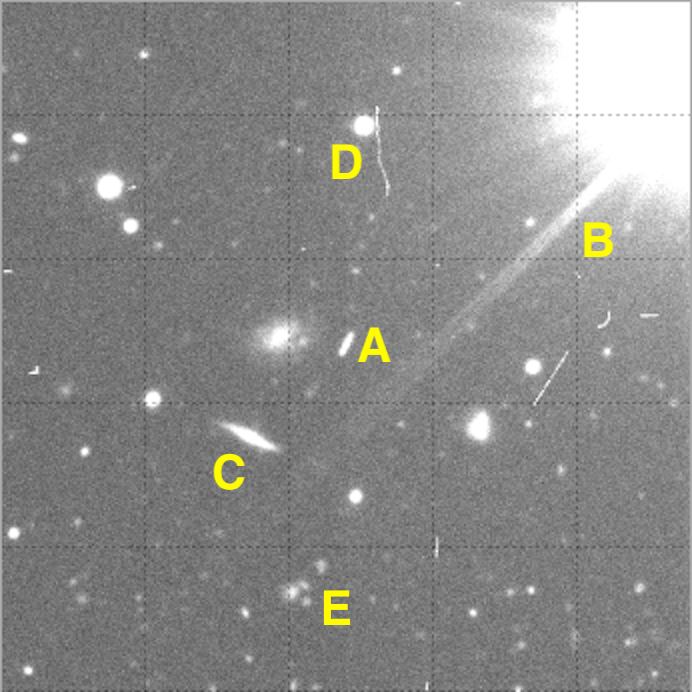}\\
	\caption{Common potential false-positives encountered in an asteroid thumbnail. \textbf{a} This thumbnail includes 4 potential false-positive sources: \textbf{A}) Asteroid (432345). \textbf{B}) \textit{Scattered light} from a bright star trails towards the lower-left corner. \textbf{C}) An \textit{extended source}, such as this edge-on galaxy, can present itself as coma if close to an asteroid. \textbf{D}) \textit{Cosmic rays} with variable morphology are common throughout our images; they can look like trails if they align with a star as in this case. \textbf{E}) \textit{Juxtaposed} objects can masquerade as active asteroids, especially when a bright object is near one or more progressively dimmer objects along the direction of apparent motion.}
	\label{FalsePositives1}
\end{figure}

We found false-positive management to be a formidable task, with specific mechanisms responsible for creating false-positives recurring throughout the project. For the rare cases where one of the authors involved in inspecting thumbnails found potential activity in an asteroid thumbnail, we checked other interpolation and color schemes, other thumbnails of the same asteroid, and the animated GIF if available. We checked frames showing the same region on the sky, including original CCD images, for background sources or image artifacts. What follows is a discussion of the primary culprits in order to convey the challenges faced during by-eye inspection (which is subjective by nature).

\textit{Juxtaposition} -- Figure \ref{FalsePositives1}A marks asteroid (432345); the object is in close proximity to a galaxy, which, if juxtaposed in a confusing manner, could give the appearance of a coma. \ref{FalsePositives1}D shows how a cosmic ray can be juxtaposed with a star. Figure \ref{FalsePositives1}E demonstrates how multiple objects may appear to be an extended source.

\textit{Extended Sources} -- Extended sources, especially galaxies, were present in a myriad of orientations and configurations. They can appear like active asteroids, as in the edge-on galaxy shown in Figure \ref{FalsePositives1}C. For a given brightness, galaxies occupied more sky area in a frame than other types of natural (i.e., non-artifact) objects and were more likely to be juxtaposed with other objects.

\textit{Scattered Light} -- Figure \ref{FalsePositives1}B is scattered light associated with an especially bright star; the flare originates from the star and tapers off the further the ``tail'' gets from the source. While obvious in Figure \ref{FalsePositives1}, the ``tail'' can be more difficult to identify as scattered light if the source is outside of the thumbnail.

\textit{Cosmic Rays} -- Cosmic rays (e.g. Figure \ref{FalsePositives1}D) are common throughout our images, most of which have exposure times of 300 seconds or longer (see Figure \ref{ExposureTimes}a). Figure \ref{FalsePositives1} D demonstrates how cosmic rays may not appear as straight lines, and they may seem to connect two or more objects together.

\textit{Poor Seeing} -- Images with poor and/or rapidly varying seeing conditions suffered from fuzziness (potentially coma-like) and elongation implying a trailed object (e.g., an asteroid).

\subsection{Limitations of By-Eye Inspection}
\label{HumanLimitations}

As proof-of-concept for future projects making use of larger datasets, we sought a general understanding of our throughput as thumbnail inspectors. It is worth noting we did not impose time limits upon ourselves. We noted markedly different inspection rates, with the time required to inspect all thumbnails ranging from 2 to 6 hours. Furthermore, our attention spans varied, with inspection sessions lasting roughly between 10 minutes to 3 hours before requiring a break. The false positive handling described above undoubtedly impacted our image examination efficacy to some degree. Given these challenges, it is evident a computational approach to screen for potential active asteroids (through e.g., PSF comparison) would improve our detection rate.

\subsection{Asteroid Selection}
\label{AsteroidSelection}
We examined only known asteroids during this work, but certainly many unknown asteroids are present within our data. Future efforts involving Citizen Scientists could locate these objects and quantify previously unrecognized biases inherent to locating activity among known asteroids. We used observations from a southern observatory, and while there may be little to no effect on observed activity occurrence rates, we acknowledge this selection effect nonetheless.

\subsection{Future Work}
\label{FutureWork}

A broader study of the efficacy of human inspectors should be carried out if employing a larger number of inspectors. Injecting artificial active asteroids into the datasets would enable quantifying detection rates. The enormous datasets (2M+ thumbnails) we plan to generate will necessitate the deployment of a Citizen Science project, an endeavor that would thoroughly flush out these detection rates.

Citizen Science endeavors enable scientists to analyze otherwise prohibitively large datasets, with the added benefit of providing the scientific community with invaluable outreach opportunities proven to engage the public and spark far-reaching interest in science. \textit{Zooniverse}\footnote{\url{www.zooniverse.org}}, designed with the average scientist in mind, facilitates deployment of crowd-sourcing science projects. Volunteers are enlisted to interpret data too complex for machines, but accomplishable by anyone with minimal training. \textit{Zooniverse} has a proven track record, with notable successes such as \textit{Galaxy Zoo} which, within 24 hours of launch, reached 70K identifications/hour \citep{Cox:2015wy}. While traditional and social media coverage undoubtedly boosted the performance of \textit{Galaxy Zoo} and other exemplary Citizen Science projects, the platform is designed to facilitate such exposure, especially through social media connectivity.

Our aim is to expand our survey to a second, comparably-sized dataset already in-hand. We will first explore strategies to quantify active asteroid candidacy through computational techniques such as PSF comparison. We will then use the combined datasets to design, implement and test a Citizen Science project. We plan to start with a moderate ($\sim$ 10 member) group of thumbnail inspectors consisting of undergraduate and graduate students, whose feedback will inform the documentation and training system which is crucial to the success of a Citizen Science project. We subsequently intend to expand our dataset to the entire DECam public archive, at which point we would open our analysis system to public participation. We hope to incorporate machine learning into our pipeline as a means of reducing the number of thumbnails sent to the Citizen Science project or to help locate candidates missed at any point in the process.

\section{Summary}
\label{summary}

We have developed an approach for finding active asteroids, rare objects visually like comets but dynamically like asteroids. We show DECam data are suitable for active asteroid searches. The approach involved processing \fitscount{} FITS files and extracting \allthumbs{} asteroid thumbnails (small images centered on an asteroid) consisting of \uniquethumbs{} unique objects. Upon visual examination of all thumbnails, we identified one previously known active asteroid (62412); our discovery rate of 1 in \uniquethumbs{} is consistent with the currently accepted active asteroid occurrence rate of 1 in 10,000. We did observe (1) Ceres and (779) Nina, though the former is a special case of \textit{a priori} activity knowledge \citep{AHearn:1992id,Kuppers:2014dp}, and neither object has ever shown signs of activity visible from Earth; as we did not observe activity in either object, we did not include them in our activity occurrence rate estimate. From our proof-of-concept study, we conclude a significantly larger survey should be carried out to locate active asteroids, finally placing them into a regime where they may be studied as a population.

\section{Acknowledgements}
\label{acknowledgements} 

 The authors thank the referee, Henry Hsieh (Planetary Science Institute), whose thorough and thoughtful feedback vastly improved the quality of this work. Prof. Ty Robinson of Northern Arizona University (NAU) helped us keep this project a priority and provided fresh perspectives on scientific dilemmas. Dr. Mark Jesus Mendoza Magbanua (University of California San Francisco) whose insights greatly improved the quality of the paper. Annika Gustaffson (NAU) provided frequent input and encouraged us to move this project forward. The enthusiastic support provided by Monsoon supercomputer system administrator Christopher Coffey (NAU) was essential in overcoming countless technical challenges. The Trilling Research Group (NAU) provided insight and feedback about our data visualization techniques. Prof. Mike Gowanlock (NAU) inspired numerous computational techniques which reduced our analysis compute time.

Computational analyses were run on Northern Arizona University's Monsoon computing cluster, funded by Arizona's Technology and Research Initiative Fund. This work was made possible in part through the State of Arizona Technology and Research Initiative Program. Michael Mommert was supported in part by NASA grant NNX15AE90G to David E. Trilling.

This research has made use of the VizieR catalogue access tool, CDS, Strasbourg, France. The original description of the VizieR service was published in A\&AS 143, 23 \citep{Ochsenbein:2000tjb}. This research has made use of data and/or services provided by the International Astronomical Union's Minor Planet Center. This research has made use of NASA's Astrophysics Data System. This research has made use of the The Institut de M\'ecanique C\'eleste et de Calcul des \'Eph\'em\'erides (IMCCE) SkyBoT Virtual Observatory tool \citep{Berthier:2006tn}. This work made use of the {FTOOLS} software package hosted by the NASA Goddard Flight Center High Energy Astrophysics Science Archive Research Center.

This project used data obtained with the Dark Energy Camera (DECam), which was constructed by the Dark Energy Survey (DES) collaboration. Funding for the DES Projects has been provided by the U.S. Department of Energy, the U.S. National Science Foundation, the Ministry of Science and Education of Spain, the Science and Technology Facilities Council of the United Kingdom, the Higher Education Funding Council for England, the National Center for Supercomputing Applications at the University of Illinois at Urbana-Champaign, the Kavli Institute of Cosmological Physics at the University of Chicago, Center for Cosmology and Astro-Particle Physics at the Ohio State University, the Mitchell Institute for Fundamental Physics and Astronomy at Texas A\&M University, Financiadora de Estudos e Projetos, Funda\c{c}\~{a}o Carlos Chagas Filho de Amparo, Financiadora de Estudos e Projetos, Funda\c{c}\~ao Carlos Chagas Filho de Amparo \`{a} Pesquisa do Estado do Rio de Janeiro, Conselho Nacional de Desenvolvimento Cient\'{i}fico e Tecnol\'{o}gico and the Minist\'{e}rio da Ci\^{e}ncia, Tecnologia e Inova\c{c}\~{a}o, the Deutsche Forschungsgemeinschaft and the Collaborating Institutions in the Dark Energy Survey. The Collaborating Institutions are Argonne National Laboratory, the University of California at Santa Cruz, the University of Cambridge, Centro de Investigaciones En\'{e}rgeticas, Medioambientales y Tecnol\'{o}gicas–Madrid, the University of Chicago, University College London, the DES-Brazil Consortium, the University of Edinburgh, the Eidgen\"ossische Technische Hochschule (ETH) Z\"urich, Fermi National Accelerator Laboratory, the University of Illinois at Urbana-Champaign, the Institut de Ci\`{e}ncies de l'Espai (IEEC/CSIC), the Institut de Física d'Altes Energies, Lawrence Berkeley National Laboratory, the Ludwig-Maximilians Universit\"{a}t M\"{u}nchen and the associated Excellence Cluster Universe, the University of Michigan, the National Optical Astronomy Observatory, the University of Nottingham, the Ohio State University, the University of Pennsylvania, the University of Portsmouth, SLAC National Accelerator Laboratory, Stanford University, the University of Sussex, and Texas A\&M University.

Based on observations at Cerro Tololo Inter-American Observatory, National Optical Astronomy Observatory (NOAO Prop. IDs 2015A-0351
2016B-0288, 2017A-0367, 2015B-0265, 2013B-0453, 2014B-0303, 2016A-0401 and 2014A-0479; PI: Scott Sheppard), which is operated by the Association of Universities for Research in Astronomy (AURA) under a cooperative agreement with the National Science Foundation.


\clearpage
\appendix

\section{Object-Specific References}
\label{ObjectReferences}

SPK-ID are found at the JPL Horizons Small Bodies Database (\url{https://ssd.jpl.nasa.gov/sbdb.cgi}).

\begin{itemize}
\setlength\itemsep{0.01mm}
	\item[\ref{Ceres}.] (1) Ceres, 1943 XB, A899 OF, SPK-ID=2000001; Activity Discovered:\cite{AHearn:1992id,Kuppers:2014dp}; Mechanism: \cite{Kuppers:2014dp}; Activity Obs.: 1 (1992) -- \cite{AHearn:1992id}, 2 (2011-2013) -- \cite{Kuppers:2014dp, Nathues:2015bx}, 3 (2015-2016) -- \cite{Thangjam:2016ho,Nathues:2017cp,Landis:2017ci,Roth:2018ki}; Visit: Dawn \citep{Russell:2016dj}; Absence of Family Association: \cite{Rivkin:2014be,Hsieh:2018kk}; Additional: \cite{Tu:2014im, Witze:2015hq, Hayne:2015cv, Nathues:2015bx, Li:2016ft, Roth:2016gr, Prettyman:2017ju, McKay:2017hb, Nathues:2017im, Landis:2017ci}

	\item[\ref{Adeona}.] (145) Adeona, SPK-ID=2000145; Activity Discovery: \cite{Busarev:2016fta}; Mechanism: \cite{Busarev:2016fta}; Activity Obs.: 1 (2012) -- \cite{Busarev:2016fta}$^\ast$; Visit: Dawn (cancelled)\footnote{\url{https://www.nasa.gov/feature/new-horizons-receives-mission-extension-to-kuiper-belt-dawn-to-remain-at-ceres}}; Additional: \cite{Busarev:2018fk}

	\item[\ref{Constantia}.] (315) Constantia, SPK-ID=2000315; Candidacy: \cite{Cikota:2014ds}; Flora family association: \cite{Alfven:1969fk}

	\item[\ref{Griseldis}.] (493) Griseldis, 1902 JS, A915 BB, SPK-ID=2000493;  Activity Discovery: \cite{Tholen:2015tu};  Activity Obs.: 1 (2015) -- \cite{Tholen:2015tu,Seargent:2017uk}; Unknown impactor size: \cite{Hui:2017et}; Absence of Family Association: \cite{Hsieh:2018kk}

	\item[\ref{Scheila}.] (596) Scheila, 1906 UA, 1949 WT, SPK-ID=2000596;  Activity Discovery: \cite{Larson:2010tt};  Mechanism: \cite{Jewitt:2011br, Bodewits:2011fo, Yang:2011it, Moreno:2011ki, Ishiguro:2011hc, Ishiguro:2011ku, Hsieh:2012kb, Husarik:2012wc, Neslusan:2016kg};  Activity Obs.: 1 (2010-2011) -- \cite{Jewitt:2011br, Bodewits:2011fo, Yang:2011it, Ishiguro:2011hc, Hsieh:2012kb, Husarik:2012wc,Neslusan:2016kg}; Absence of Family Association: \cite{Hsieh:2018kk}

	\item[\ref{Interamnia}.] (704) Interamnia, 1910 KU, 1952 MW, SPK-ID=2000704; Activity Discovery, Mechanism: \cite{Busarev:2016fta}; Activity Obs.:  1 (2012) -- \cite{Busarev:2016fta}$^\ast$; Absence of Family Association: \cite{Rivkin:2014be}; Shape Model: \cite{Sato:2014be}; Additional: \cite{Busarev:2018fk}

	\item[\ref{Nina}.] (779) Nina, 1914 UB, A908 YB, A912 TE, SPK-ID=2000779;  Activity Discovery, Mechanism: \cite{Busarev:2016fta}; Activity Obs.:  1 (2012) -- \cite{Busarev:2016fta}$^\ast$, 2 (2016) -- \cite{Busarev:2018fk}

	\item[\ref{Ingrid}.] (1026) Ingrid, 1923 NY, 1957 UC, 1963 GD, 1981 WL8, 1986 CG2, 1986 ES2, SPK-ID=2001026; Candidacy: \cite{Cikota:2014ds}; Follow-up Observation (negative): \cite{Betzler:2015uy}; Flora family association: \cite{Alfven:1969fk}; Additional: \cite{Nakano:1986vl,Busarev:2018fk}

	\item[\ref{Beira}.] (1474) Beira, 1935 QY, 1950 DQ, SPK-ID=2001474;  Activity Discovery: \cite{Busarev:2016fta};  Mechanism: \cite{Busarev:2016fta};  Activity Obs.:  1 (2012) -- \cite{Busarev:2016fta}$^\ast$; Chaotic Cometary Orbit: \cite{Hahn:1985ds}; Additional: \cite{Busarev:2018fk}

	\item[\ref{Oljato}.] (2201) Oljato, 1947 XC, 1979 VU2, 1979 XA, SPK-ID=2002201;  Activity Discovery: \cite{Russell:1984eq}; Activity Obs.: 1 (1984) -- \cite{Russell:1984eq}, Negative (1996) -- (\cite{Chamberlin:1996gx};  Visit: \cite{Perozzi:2001gp}; Additional: \cite{Kerr:1985gp, McFadden:1993cl, Connors:2016ge}

	\item[\ref{Phaethon}.] (3200) Phaethon, 1983 TB, SPK-ID=2003200; Activity Discovery: \cite{Battams:2009tt};  Mechanism: ;  Activity Obs.: Negative -- \cite{Chamberlin:1996gx,Hsieh:2005ev}, 1 (2009) --  \cite{Battams:2009tt,Jewitt:2010jd} 2 (2012) -- \cite{Li:2013fs,Jewitt:2013jm}, 3 (2016) -- (\cite{Hui:2017kz};  Visit: Destiny Plus \citep{IWATA:2016gua};  Pallas Family Association: \cite{Todorovic:2018dz}; Additional: \cite{Jewitt:2010jd, Ryabova:2012bo, Li:2013fs, Jewitt:2013jm, Ansdell:2014hs, Jakubik:2015dj, Hanus:2016ez, Sarli:2017ey}

	\item[\ref{DonQuixote}.] (3552) Don Quixote, 1983 SA, SPK-ID=2003552; Activity Discovery, Mechanism: \cite{Mommert:2014gl}; Activity Obs.: 1 (2009) -- \cite{Mommert:2014gl}, (2018) -- \cite{Mommert:2018vl}; Chaotic Cometary Orbit (as 1983 SA): \cite{Hahn:1985ds}

	\item[\ref{Aduatiques}.] (3646) Aduatiques, 1985 RK4, 1979 JL, 1981 WZ6, SPK-ID=2003646; Candidacy: \cite{Cikota:2014ds}; Follow-up (inconclusive): \cite{SosaOyarzabal:2014vw}

	\item[\ref{WilHar}.] (4015) Wilson-Harrington, 1979 VA, 107P,  SPK-ID=2004015; Activity Discovery: \cite{Cunningham:1950wt};  Activity Obs.: 1 (1949) -- \cite{Cunningham:1950wt}, 2 (1979) -- \cite{Degewij:1980uk}, Negative (1992) -- \cite{Bowell:1992tp}, Negative (1996) \cite{Chamberlin:1996gx}, Negative (2008) -- \cite{Licandro:2009bj}, Negative (2009-2010) -- \cite{Ishiguro:2011hm,Urakawa:2011en}, 3-6 (1992, 1996, 2008, 2009-2010) \cite{Ferrin:2012gm};  Visits: Failed \citep{Rayman:2001gh}, Concept \citep{Sollitt:2009vt};  Chaotic Cometary Orbit (as 1979 VA): \cite{Hahn:1985ds}; Additional: \cite{Harris:1950ub, vanBiesbroeck:1951ti, Helin:1980wo, Helin:1981tx, Osip:1995dr, Fernandez:1997gb}

	\item[\ref{24684}.] (24684), 1990 EU4, 1981 UG28, SPK-ID=2024684; Candidacy: \cite{Cikota:2014ds}

	\item[\ref{35101}.] (35101) 1991 PL16, 1998 FZ37, SPK-ID=2035101; Candidacy: \cite{Cikota:2014ds}; Eunomia Family Association: \cite{Cikota:2014ds}

	\item[\ref{62412}.] (62412), 2000 SY178, SPK-ID=2062412;  Activity Discovery: \cite{Sheppard:2015cw}; Activity Obs.: 1 (2014) \cite{Sheppard:2015cw}; Hygiea Family Association: \cite{Sheppard:2015cw,Hsieh:2018kk}

	\item[\ref{Ryugu}.] (162173) Ryugu, SPK-ID=2162173;  Activity Discovery, Mechanism, Activity Obs.: 1 (2007) -- \cite{Busarev:2018fk}$^\ast$;  Visit: Hayabusa 2 \citep{Tsuda:2013hc}; Clarissa Family Association \cite{Campins:2013bp,LeCorre:2018hl}; Thermal Inertia: \cite{Liangliang:2014hs}; Additional: \cite{Suzuki:2018if, Perna:2017bz}

	\item[\ref{GO98}.] (457175), 2008 GO98, 362P, SPK-ID=2457175;  Activity Discovery: \cite{Kim:2017cw};  Activity Obs.: 1 (2017) \cite{Masi:2017ty};  Hilda Family Association: \cite{Warner:2018tu}; Additional: \cite{Sato:2017ty, Yoshimoto:2017ty, Birtwhistle:2017ty, Bacci:2017ty, Bell:2017ty, Bryssinck:2017ty}

	\item[\ref{ElstPizarro}.] 133P/Elst-Pizarro, (6968), 1979 OW7, 1996 N2, SPK-ID=2007968;  Activity Discovery: \cite{Elst:1996vj};  Mechanism: \cite{Hsieh:2004gd, Jewitt:2014gp};  Activity Obs.: 1 (1996) \cite{Elst:1996vj}, 2 (2002) \cite{Hsieh:2004gd}, Negative (2005) \cite{Toth:2006jw}, 2 (2007) \cite{Hsieh:2010fm, Bagnulo:2010fm, Rousselot:2011kk}, 3 (2013) \cite{Jewitt:2014gp};  Visit: Castalia \citep{Snodgrass:2017bi};  Themis Family Association: \cite{Boehnhardt:1998wr}; Additional: \cite{Toth:2000wm, Ferrin:2006gr, Prialnik:2009ga}

	\item[\ref{176P}.] 176P/LINEAR, (118401), P/1999 RE$_{70}$, 2001 AR7, SPK-ID=2118401;  Activity Discovery: \cite{Hsieh:2006tj,Hsieh:2009gy};  Mechanism: \cite{Hsieh:2014cl}; Activity Obs.: 1 (2005) \cite{Hsieh:2006tj}, Negative (2006-2009) \cite{Hsieh:2011fw}, Negative (2011) \cite{Hsieh:2014cl}; Themis Family Association: \cite{Hsieh:2009gy,Hsieh:2018kk} Additional: \cite{Hsieh:2009bl,Licandro:2011jh,deValBorro:2012iy}

	\item[\ref{233P}.] 233P (La Sagra), P/2009 W$_{50}$, 2005 JR71, SPK-ID=1003062; Activity Discovery: \cite{Mainzer:2010uo}, Activity Obs.: 1 (2009) \cite{Mainzer:2010uo}; Absence of Family Association: \cite{Hsieh:2018kk}

	\item[\ref{238P}.] 238P/Read, P/2005 U1, 2010 N2, SPK-ID=1001676;  Activity Discovery: \cite{Read:2005to};  Activity Obs.: 1 (2005) \cite{Read:2005to}, 2 (2010) \cite{Hsieh:2011kl}, 3 (2016) \cite{Hsieh:2017tj}; Gorchakov Family Association: \cite{Hsieh:2018kk}; Former Themis Family Association: \cite{Haghighipour:2009kl}; Additional: \cite{Hsieh:2009ek, Pittichova:2010uc}

	\item[\ref{259P}.] 259P/Garradd, 2008 R1, SPK-ID=1002991; Activity Discovery: \cite{Garradd:2008ts}; Mechanism: \cite{Jewitt:2009jb}; Activity Obs.: 1 (2008) \cite{Garradd:2008ts}, 2 (2017) \cite{Hsieh:2017wg,Hsieh:2017tj}; Absence of Family Association: \cite{Hsieh:2018kk}; Additional: \cite{Kossacki:2012du, MacLennan:2012cj, Kleyna:2012hh}

	\item[\ref{288P}.] 288P, (300163), 2006 VW139, SPK-ID=2300163;  Activity Discovery: \cite{Hsieh:2011wj}; Activity Obs.: 1 (2011) \cite{Hsieh:2011wj}, 2 (2016-2017) \cite{Agarwal:2017cj,Hsieh:2017tj}; Themis Family Association: \cite{Hsieh:2012he,Hsieh:2018kk}; Additional: \cite{Hsieh:2012he, Novakovic:2012bf, Agarwal:2016dc}

	\item[\ref{311P}.] 311P/Pan-STARRS, P/2013 P5, SPK-ID=1003273; Activity Discovery: \cite{Micheli:2013tj};  Mechanism: \cite{Jewitt:2013id, Moreno:2014gf, Hainaut:2014im, Jewitt:2015br};  Activity Obs.: 1 (2013-2014) \cite{Micheli:2013tj,Jewitt:2015br}; Behrens Family Association: \cite{Hsieh:2018kk}

	\item[\ref{313P}.] 313P/Gibbs, P/2014 S4, 2003 S10, SPK-ID=1003344;  Activity Discovery: \cite{Gibbs:2014ww};  Mechanism, Activity Obs.: 1 (2003) \cite{Nakano:2014un,Skiff:2014wx,Hui:2015bu}, 2 (2015) \cite{Jewitt:2015iq}; Lixiaohua Family Association: \cite{Hsieh:2013ie,Hsieh:2015bc,Hsieh:2018kk}; Additional: \cite{Jewitt:2015gb, Hsieh:2015bc, Pozuelos:2015hg}

	\item[\ref{324P}.] 324P/La Sagra, P/2010 R2, 2015 K3, SPK-ID=1003104; Activity Discovery: \cite{Nomen:2010ut};  Activity Obs.: 1 (2010-2011) \cite{Nomen:2010ut,Hsieh:2012gd}, Negative (2013) \cite{Hsieh:2014bg}, 2 (2015) \cite{Hsieh:2015jv,Jewitt:2016hw}; Alauda Family Association: \cite{Hsieh:2018kk};  Additional: \cite{Moreno:2011bh,Hsieh:2012gd,Hsieh:2014bg,Hsieh:2015jv}

	\item[\ref{331P}.] 331P/Gibbs, P/2012 F5, SPK-ID=1003182;  Activity Discovery: \cite{Gibbs:2012uo};  Mechanism: \cite{Stevenson:2012dk, Drahus:2015gm};  Activity Obs.: 1 (2012) \cite{Gibbs:2012uo}, 2 (2015) \cite{Drahus:2015gm}; Gibbs family association: \cite{Novakovic:2014eb}; Additional: \citep{Stevenson:2012dk,Moreno:2012kl}

	\item[\ref{348P}.] 348P, P/2017 A2, P/2011 A5 (PANSTARRS), SPK-ID=1003492; Activity Discovery: \cite{Wainscoat:2017vd}; Activity Obs.: 1 (2017) \cite{Wainscoat:2017vd}; Absence of Family Association: \cite{Hsieh:2018kk}

	\item[\ref{354P}.] 354P/LINEAR, P/2010 A2, 2017 B5, SPK-ID=1003055;  Activity Discovery: \cite{Birtwhistle:2010us};  Activity Obs.: 1 (2010) \cite{Birtwhistle:2010us,Jewitt:2010us}; Baptistina Family Association: \cite{Hsieh:2018kk}; Additional: \cite{Moreno:2010fb,Jewitt:2010bs,Snodgrass:2010gl,Jewitt:2011cw,Hainaut:2012ee,Kim:2012kk,Agarwal:2012vz,Kleyna:2013ge,Jewitt:2013di,Agarwal:2013dv,Kim:2017cw,Kim:2017di}

	\item[\ref{358P}.] 358P/PanSTARRS, P/2012 T1, 2017 O3, SPK-ID=1003208;  Activity Discovery: \cite{Wainscoat:2012tk}; Activity Obs.: 1 (2012) \cite{Wainscoat:2012tk}, 2 (2017) \cite{Kim:2017cw};  Mechanism: \cite{Hsieh:2013ie}; Lixiaohua Family Association: \citep{Hsieh:2013ie,Hsieh:2018kk}; Additional: \cite{Moreno:2013jn, ORourke:2013ca, Snodgrass:2017jg}

	\item[\ref{P2013R3}.] P/2013 R3 (Catalina-Pan-STARRS), SPK-ID=1003275 (P/2013 R3-A SPK-ID=1003333, P/2013 R3-B SPK-ID=1003334); Activity Discovery: \cite{Bolin:2013wp,Hill:2013wp}; Activity Obs.: 1 (2013-2015) \cite{Bolin:2013wp,Hill:2013wp,Jewitt:2017fa}; Mandragora Family Association: \cite{Hsieh:2018kk}; Additional: \cite{Jewitt:2014fe, Hirabayashi:2014de}

	\item[\ref{P2015X6}.] P/2015 X6 (Pan-STARRS), SPK-ID=1003426;  Activity Discovery: \cite{Lilly:2015vg};  Activity Obs.: 1 (2015) \cite{Lilly:2015vg,Tubbiolo:2015tq,Moreno:2016bi}; Aeolia Family Association: \cite{Hsieh:2018kk}

	\item[\ref{P2016G1}.] P/2016 G1 (Pan-STARRS), SPK-ID=1003460;  Activity Discovery: \cite{Weryk:2016tf};  Mechanism: \cite{Moreno:2016hy}; Activity Obs.: 1 (2016) \cite{Weryk:2016tf,Moreno:2017bi}; Adeona Family Association: \cite{Hsieh:2018kk}

	\item[\ref{P2016J1}.] P/2016 J1 (Pan-STARRS), P/2016 J1-A (SPK-ID=1003464), P/2016 J1-B (SPK-ID=1003465);  Activity Discovery: \cite{Wainscoat:2016vh}; Activity Obs.: 1 (2016) \cite{Wainscoat:2016vh,Hui:2017gq}; Theobalda Family Association: \cite{Hsieh:2018kk}

\end{itemize}
$^\ast$: Under review.

\bibliography{papers.bib}

\end{document}